\documentclass[twocolumn, twocolappendix]{aastex701}

\newcommand{\referee}[1]{#1}

\usepackage{caption}
\usepackage{amsmath}

\begin{document}

\title{Connecting JWST Silicate Cloud Observations to Exoplanet Cloud Microphysics with \texttt{Nimbus}}

\author[orcid=0000-0003-1285-3433,sname='Kiefer']{Sven Kiefer}
\affiliation{Department of Astronomy, University of Texas at Austin, 2515 Speedway, Austin, TX 78712, USA}
\email[show]{sven.kiefer@utexas.edu}  

\author[orcid=0000-0002-4404-0456,sname='Morley']{Caroline V. Morley} 
\affiliation{Department of Astronomy, University of Texas at Austin, 2515 Speedway, Austin, TX 78712, USA}
\email{cmorley@utexas.edu}  

\author[orcid=0000-0003-4225-6314,sname='Rowland']{Melanie Rowland}
\affiliation{Department of Astronomy, University of Texas at Austin, 2515 Speedway, Austin, TX 78712, USA}
\affiliation{Department of Astrophysics, American Museum of Natural History, New York, NY 10024, USA}
\email{sven.kiefer@utexas.edu}

\begin{abstract}

The unprecedented accuracy of JWST has led to the detection of silicate clouds in exoplanet atmospheres, allowing for the first time to probe cloud formation in extreme environments. While parametrized cloud descriptions can fit these observations, the results do not fully agree with microphysical models. To bridge this gap, we developed \texttt{Nimbus}, a fast microphysical cloud model that can constrain cloud formation processes from observations \referee{and utilize \texttt{Virga}, an equilibrium condensation model balancing gravitational settling and diffusion.} Using both models, we investigate WASP-107~b, WASP-17~b, VHS-1256~b, and YSES-1~c to determine their cloud structure and constrain cloud formation processes. Our results show that all four planets have cluster-sized silicate particles \referee{($r\sim1$ nm)} at high altitudes. Within \texttt{Nimbus} and \texttt{Virga}, these particles can only be explained by highly inefficient cloud particle settling ($f_\mathrm{sed} < 0.1$) or by inefficient growth rates due to low sticking coefficients ($s < 10^{-4}$). Our results also show that the sticking coefficient is directly linked to the vertical extent of clouds and can therefore be constrained using the broad shape of the spectral energy distribution. The sticking coefficients found for VHS-1256~b and YSES-1~c are in agreement with expectations from laboratory experiments under Earth-like conditions \referee{($0.01 < s < 0.3$)}. Panchromatic observations were crucial to achieve these constraints. Future cloud studies should therefore aim to combine observational data from 1~$\mu$m to 10~$\mu$m whenever possible.

\end{abstract}



\section{Introduction} 
\label{sec:Introduction}

    Cloud models for exoplanets predict that most planets with a substantial atmosphere will have clouds \citep[see e.g.][]{marley_clouds_2013,  mbarek_clouds_2016, powell_formation_2018, fauchez_impact_2019, gao_aerosol_2020, herbort_atmospheres_2020, herbort_atmospheres_2022, helling_exoplanet_2023, arfaux_physically_2023}. These predictions are supported by the high prevalence of silicate clouds in brown dwarfs of the same temperatures. Silicon absorption features can typically be observed in spectra of brown dwarfs ranging from L2 to L8 \citep{suarez_ultracool_2022}. Models of varying complexity have been developed to simulate these cloud structures \citep{ackerman_precipitating_2001, allard_limiting_2001, helling_dust_2001, helling_dust_2006, morley_neglected_2012} and to fit the observed silicate absorption features \citep{luna_empirically_2021, burningham_cloud_2021, vos_patchy_2023}. Before the launch of the \textit{James Webb Space Telescope} (JWST), evidence for clouds in exoplanet atmospheres could only be found through the observations of muted or absent molecular features \citep{bean_ground-based_2010, kreidberg_clouds_2014, espinoza_access_2019, spyratos_transmission_2021, libby-roberts_featureless_2022, lustig-yaeger_jwst_2023}. However, flat spectra can also result from non-substantial atmospheres \citep{lustig-yaeger_jwst_2023, xue_jwst_2024, wachiraphan_thermal_2025}, or from large observational errors \citep{jiang_featureless_2023, lim_atmospheric_2023}. With JWST it is now possible to confirmed the presence of silicate clouds in exoplanet atmospheres by observing the same spectral features of silicon bearing species as in brown dwarfs \citep[e.g.][]{grant_jwst-tst_2023, dyrek_so2_2023, miles_jwst_2023, inglis_quartz_2024}. Some of these observations have found Si-O bearing species at altitudes higher than clouds are expected, raising question on how clouds form in exoplanet atmospheres \citep{cushing_spitzer_2006, burningham_cloud_2021, molliere_evidence_2025}. Microphysical cloud formation models are therefore needed to investigate these new discoveries.

\begin{figure}
       \centering
       \includegraphics[width=\linewidth]{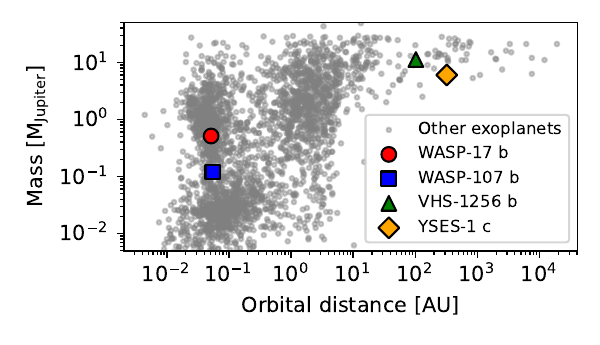}
       \caption{Orbital distance and mass of the four planets studied in this work compared to other exoplanets.}
       \label{fig:planets}
    \end{figure}

    In this paper, we analyse four exoplanets with observed cloud features in JWST data (see: Fig.~\ref{fig:planets}). WASP-107~b \citep{anderson_discoveries_2017} and WASP-17~b \citep{anderson_wasp-17b_2009} are two hot Jupiters where transit spectroscopy has revealed silicon-bearing cloud particles in the upper atmosphere \citep{dyrek_so2_2023, welbanks_high_2024, grant_jwst-tst_2023}. \citet{dyrek_so2_2023} found that a mixture of Si-bearing cloud species and carbon achieves the best fit of WASP-107~b. The Si-O bond feature observed in WASP-17~b is best explained with the presence of SiO$_2$ particles \citep{grant_jwst-tst_2023}. VHS-1256~b\footnote{Full name: VHS J125601.92-125723.9 b} \citep{gauza_discovery_2015} and YSES-1~c\footnote{Also known as TYC 8998-760-1 c} \citep{bohn_two_2020} are wide-orbit companions where thermal emission spectroscopy showed clear Si-O bond absorption features around 10~$\mu$m \citep{miles_jwst_2023, hoch_silicate_2025}. Even though both transmission and thermal emission spectroscopy detect the same Si-O bond feature, transmission spectra typically probe higher altitudes ($ p \approx 10^{-3}$~bar) than thermal emission spectra ($p \approx 10^{-1}$~bar) because of their unique viewing geometry \citep{fortney_effect_2005}. Each method therefore provides insights into different altitudes of the atmosphere. 
    
    Cloud formation depends on the chemical composition, thermal structure, and mixing efficiency of the atmospheres as well as the thermodynamic properties of the cloud forming materials \citep{helling_exoplanet_2019, gao_aerosols_2021}. Cloud models of various complexity have therefore been developed to manage the trade off between sophistication and computational feasibility (Sect.~\ref{sec:Model_Nimbus}). Retrieval models often use parametrized clouds which do not consider the physics of cloud formation. \texttt{Virga} offers an intermediate step between physical and parametrized cloud models \citep{batalha_condensation_2026}. This model is based on the \citet{ackerman_precipitating_2001} model often called `EddySed' in the literature. It derives physically informed cloud structures by assuming phase-equilibrium and a given settling efficiency of the cloud particles (see Sect.~\ref{sec:Model_Virga}). Models that include a full microphysical description of cloud formation \citep[e.g.][]{helling_dust_2006, gao_microphysics_2018, woitke_dust_2020} can simulate more realistic cloud structures but are often not suitable for retrievals due to their long computation time. Characterizing cloud microphysics from observations therefore requires a light-weight, microphysical cloud model that supports efficient parameter exploration. Here, we present the new cloud model \texttt{Nimbus} which is specifically developed to meet these requirements (Sect.~\ref{sec:Model}). We use \texttt{Virga} and \texttt{Nimbus} to constrain cloud microphysics in the transmission spectra of WASP-107~b and WASP-17~b (Sect.\ref{sec:obs_trans}) and the thermal emission spectra of VHS-1256~b, and YSES-1~c (Sect.~\ref{sec:obs_emis}). We discuss the derived constraints and trends in cloud particle properties in Sect.~\ref{sec:Discussion}, and summarize our findings in Sect.~\ref{sec:Conclusion}.

\section{Cloud Modelling}
\label{sec:Model}

    \referee{For this work, we use two cloud models: \texttt{Virga} (Sect.~\ref{sec:Model_Virga}), a versatile model for phase-equilibrium clouds, and the newly developed \texttt{Nimbus} (Sect.~\ref{sec:Model_Nimbus}), a time-dependent microphysical cloud model optimized to constrain cloud formation processes from observations.}

    \subsection{\texttt{Virga}}
    \label{sec:Model_Virga}

    \begin{figure}
        \centering
        \includegraphics[width=1\linewidth]{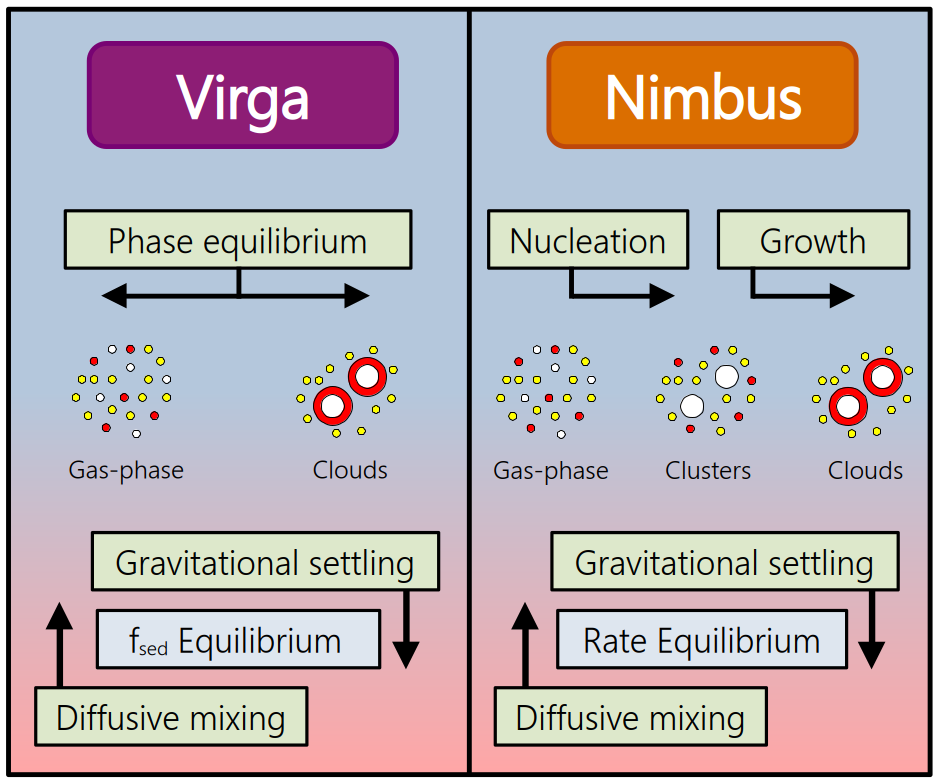}
        \caption{Comparison between \texttt{Virga} and \texttt{Nimbus} \textbf{Left:} \texttt{Virga} assumes phase equilibrium of the gas-phase and mass exchange equilibrium between diffusive upward mixing and gravitational settling. \textbf{Right:} \texttt{Nimbus} considers the rates of nucleation and accretion. Each model is run until mass exchange equilibrium between diffusive upward mixing and gravitational settling is reached.}
        \label{fig:vig_nimb_comp}
    \end{figure}
    
    \texttt{Virga} \citep{batalha_natashabatalhavirga_2020, batalha_condensation_2026, moran_fractal_2025} balances the diffusion of gas-phase species and cloud particles with the gravitational settling of cloud particles. For each cloud forming material, the following equation is solved:
    \begin{equation}
        \label{eq:aandm}
        -K_{zz} \frac{\partial q_t}{\partial z} - f_\mathrm{sed} \omega_\star q_c = 0
    \end{equation}
    \referee{where $K_{zz}$ [cm$^2$/s] is the diffusion constant (Sect.~\ref{sec:Model_mixing}),} $z$ [cm] is the altitude, $\omega_\star$ [cm/s] is the convective velocity scale, $f_\mathrm{sed}$ is the settling parameter, $q_c$~[g/g] is the Mass Mixing Ratio (MMR) of the solid cloud particles, $q_v$~[g/g] is the MMR of cloud forming material in the gas-phase, and $q_t = q_c + q_v$~[g/g] is the total MMR of the cloud material.

    The settling parameter $f_\mathrm{sed}$ describes the efficiency of upward diffusion compared to the gravitational settling of cloud particles \citep{ackerman_precipitating_2001, rooney_new_2022}:
    \begin{equation}
        \label{eq:general_fsed}
        f_\mathrm{sed}(z) = \frac{\int^\infty_0 v_\mathrm{dr}(r, z)~\chi(r)~dr}{\omega_\star (z)}
    \end{equation}
    \referee{where $r$~[cm] is the cloud particle radius, $v_\mathrm{dr}$~[cm/s] is the settling velocity (Sect.~\ref{sec:Model_vdr}), $\chi(r)$~[1/cm] is the Particle Size Distribution (PSD; Sect.~\ref{sec:Model_PSD}).} The convective velocity scale can be approximated by the eddy diffusion coefficient $K_{zz}$ and the mixing length $L$~[cm]. The mixing length scale itself can be approximated by the atmospheric scale height $H$~[cm]:
    \begin{equation}
        \omega_\star(z) \approx \frac{K_{zz}}{L} \approx \frac{K_{zz}}{H} = \frac{\mu g K_{zz} }{R_gT}
    \end{equation}
    \referee{where $\mu$~[g/mol] is the mean molecular weight of the atmosphere, $g$~[cm/s$^2$] is the gravity, $T$~[K] is the temperature of the atmosphere, and $R_g = 8.314 \times 10^7$~erg/(K~mol) is the ideal gas constant.} The integral of Eq.~\ref{eq:general_fsed} does not have an analytic solution and must be approximated numerically \citep[see e.g.][]{ackerman_precipitating_2001}.

    The settling parameter $f_\mathrm{sed}$ links the settling velocity $v_\mathrm{dr}(r, z)$ to the diffusive mixing and therefore allows to determine the cloud particle radius from the atmospheric structure. Some studies have assumed a constant $f_\mathrm{sed}(z) \equiv f_0$ throughout the atmosphere and have used $f_0$ as a hyper-parameter \citep[see e.g.][]{robbins-blanch_cloudy_2022, christie_impact_2022, inglis_quartz_2024, mukherjee_cloudy_2025}. To allow for a height dependent $f_\mathrm{sed}$, \cite{rooney_new_2022} have derived the following parametrisation:
    \begin{equation}
        f_\mathrm{sed}(z) = \alpha_f \exp \left( \frac{z - z_T}{6 \beta_f H_0} \right) + \epsilon_f
    \end{equation}
    where $\alpha_f$, $\beta_f$, and $\epsilon_f$ are free parameters, $H_0$ [cm] and $z_T$ [cm] is the reference scale height and reference altitude, respectively. For this paper, we have further reduced the constraints on $f_\mathrm{sed}$ by allowing a free height dependence. This allows to account for height dependent changes in the cloud microphysics, e.g., the nucleation and growth rate.

    The \texttt{Virga} cloud structures also depend on the assumed MMRs of the cloud forming species. The standard MMRs for \texttt{Virga} at solar metallicity are:
    \begin{align}
        \label{eq:mmr_sio}
        &\mathrm{MMR_\mathrm{SiO}} = 1.14 \cdot 10^{-3} \\
        \label{eq:mmr_sio2}
        &\mathrm{MMR_\mathrm{SiO_2}} = 1.55 \cdot 10^{-3} \\
        \label{eq:mmr_mgsio3}
        &\mathrm{MMR_\mathrm{MgSiO_3}} = 2.59 \cdot 10^{-3} \\
        \label{eq:mmr_mg2sio4}
        &\mathrm{MMR_\mathrm{Mg_2SiO_4}} = 3.57 \cdot 10^{-3}
    \end{align}
    To achieve a better fit to observations, we later allow the cloud material MMR to vary and constrain it through $\chi^2$-minimisation.

    \subsection{\texttt{Nimbus}}
    \label{sec:Model_Nimbus}

    \begin{table*}
        \centering
        \caption{\referee{Microphysical parameters of cloud particle materials. All radii are approximated by the volume and density of the species. Sources: (1) \citet{lee_modelling_2023} (2) \citet{lee_dust_2018} (3) \citet{gail_seed_2013}, (4) \citet{gail_primary_1986}, (5) \citet{grant_jwst-tst_2023}, (6) \citet{woitke_ggchem_2021}, (7) \citet{ackerman_precipitating_2001}, (8) \citet{kozasa_formation_1989}, (9) \citet{sindel_revisiting_2022}, (10) \citet{kozasa_grain_1987}, and (11) \citet{pradhan_growth_2003}.}
        }
        \label{tab:microparam}
        \begin{tabular}{l l l l l l l l}
            \hline\hline
            Species       & $\rho_c$ [g/cm$^{3}$] & $r_n$ [cm]               & $\mu_c$ [g/mol] & $\sigma$~[erg/cm$^{2}$] & $p_\mathrm{vap}$~[dyn/cm$^{2}$]                   & Source \\
            \hline
            SiO           & 2.18                  & $2.001 \times 10^{-8}$   & 44.085          & $500$                   & $\exp(-49520/T + 32.52)$                          & 1, 2, 3, 4\\
            SiO$_2$       & 2.65                  & $2.079 \times 10^{-8}$   & 60.084          & $243.2 + 0.031~T$       & $10^{- 28265/T + 13.168 + 6}$                     & 1, 5, 6 \\
            MgSiO$_3$     & 3.19                  & $2.319 \times 10^{-8}$   & 100.39          & $197.3 + 0.098~T$       & $10^{6} \exp(-58663/T + 25.37)$                   & 1, 7\\
            Mg$_2$SiO$_4$ & 3.21                  & $2.590 \times 10^{-8}$   & 140.69          & $436$                   & $10^{6} \exp(-62279/T + 20.94)$                   & 1, 8\\
            TiO$_2$       & 4.23                  & $1.956 \times 10^{-8}$   & 79.866          & $589.8 - 0.071~T$       & $\exp(-77044/T + 40.31 - \textnormal{2.59e-3}~T$  & 1, 2, 6, 9\\
                          &                       &                          &                 &                         & $\quad \quad + \textnormal{6.02e-7}~T^2 - \textnormal{6.87e-11}~T^3)$  & \\
            Fe            & 7.87                  & $1.412 \times 10^{-8}$   & 55.845          & $2565 - 0.39~T$         & $10^{6} \exp(-47664/T + 15.71)$                   & 1, 7\\
                          &                       &                          &                 &                         & $\quad \times 10^{- \textnormal{2.84e-7}~T^2 + \textnormal{1.83e-10}~T^3}$ & \\
            Al$_2$O$_3$   & 3.96                  & $2.169 \times 10^{-8}$   & 101.96          & $1024 - 0.177~T$        & $10^{6} \exp(-73503/T + 22.01)$                   & 1, 10, 11 \\
            \hline
        \end{tabular}
    \end{table*}

    \begin{table*}
        \centering
        \caption{Configuration of different micro-physical cloud models. \referee{The abbreviations are gas-phase mixing scheme (G. Mix.),} diffusion (Dif.), advection (Adv.), parametrized (Param.), collisonal (Col.), morphology (Morph.), heterogenous (Hetero.), Homogenous (Homo.), mono-dispersed (Mono.), log-normal (logn.), coagulation (Coa.), high Knudsen number $Kn^\mathrm{high}$ ($K^+$), and low Knudsen number $Kn^\mathrm{low}$ ($K^-$). All models except \citet{lee_modelling_2023} are one-dimensional.}
        \label{tab:cloud_models}
        \begin{tabular}{l l l l l l l l l l}
            \hline\hline
             Name               & Model                             & G. Mix.     & $J$   & $G$          & Morph.  & PSD     & Coa.      & $K^+$ & $K^-$ \\  
            \hline 
            \texttt{Drift} & \citet{helling_dust_2006}    & Relaxation & MCNT   & Col.\&Dif.  & Hetero. & Mono.   & x          & \checkmark & \checkmark \\
            Unnamed             & \citet{ohno_microphysical_2018}   & Diffusion  & Param. & Col.\&Dif.  & Homo.   & Mono.   & \checkmark & \checkmark & \checkmark \\
            \texttt{CARMA}      & \citet{gao_microphysics_2018}     & Diffusion  & MCNT   & Diffusive   & Homo.   & Binned  & \checkmark & x          & \checkmark \\
            \texttt{ARCiS}      & \citet{ormel_arcis_2019}          & Diffusion  & Param. & Collisional & Homo.   & Mono.   & \checkmark & \checkmark & x          \\
            \texttt{DiffuDrift} & \citet{woitke_dust_2020}          & Diffusion  & MCNT   & Collisional & Hetero. & Mono.   & x          & \checkmark & x          \\
            \texttt{Mini-cloud} & \citet{lee_modelling_2023}        & Dif.\&Adv. & MCNT   & Col.\&Dif.  & Hetero. & Various & \checkmark & \checkmark & \checkmark \\
            \texttt{ExoLyn}     & \citet{huang_exolyn_2024}         & Diffusion  & Param. & Collisional & Hetero. & Mono.   & \checkmark & \checkmark & \checkmark \\
            \texttt{Virga}      & \citet{batalha_condensation_2026} & Diffusion  & -      & -           & Homo. & Logn. & x & \checkmark & \checkmark  \\
            \texttt{Nimbus}     & This paper                        & Diffusion  & MCNT   & Col.\&Dif.  & Homo.   & Mono.   & x          & \checkmark & \checkmark \\
            \hline
        \end{tabular}
    \end{table*}

    \referee{\texttt{Nimbus} is a time-dependent, one-dimensional cloud formation model. It considers gas-phase transport through diffusion (Sect.~\ref{sec:Model_mixing}), cloud particle transport through diffusion and gravitational settling (Sect.~\ref{sec:Model_vdr}), nucleation through Modified Classical Nucleation Theory (MCNT; Sect.~\ref{sec:Model_nuc}), and growth of material in the diffusion and collision limited regime (Sect.~\ref{sec:Model_acc}). Each cloud material is assumed to nucleate and grow separately by solving the following set of equations:}
    \begin{align}
        \label{eq:nimb_1}
        \rho_a \frac{d q_c}{dt} &= \frac{\partial}{\partial z} q_c \rho_a v_\mathrm{dr} - \frac{\partial}{\partial z} K_{zz} \rho_a \frac{\partial q_c}{\partial z} + m_\mathrm{n} J + m_\mathrm{1} G \\
        \label{eq:nimb_2}
        \rho_a \frac{d q_n}{dt} &= \frac{\partial}{\partial z} q_n \rho_a v_\mathrm{dr} - \frac{\partial}{\partial z} K_{zz} \rho_a \frac{\partial q_n}{\partial z} + m_\mathrm{n} J \\
        \label{eq:nimb_3}
        \rho_a \frac{d q_v}{dt} &= -\frac{\partial}{\partial z} K_{zz} \rho_a \frac{\partial q_v}{\partial z} - m_\mathrm{n} J - m_\mathrm{1} G
    \end{align}
    \referee{where $\rho_a$~[g/cm$^{3}$] is the atmospheric density, $q_n$~[g/g] is the MMR of the CCN, $m_\mathrm{n}$~[g] is the CCN mass, $m_1$~[g] is the mass of a single unit of a cloud forming material, $J$~[1/cm$^{3}$/s] is the nucleation rate, and $G$~[1/cm$^{3}$/s] is the growth rate. From these parameters, the cloud particle number density $n_c$~[1/cm$^{3}$], the cloud particle radius $r$, and the mean cloud particle mass $m_\mathrm{avg}$ [g] can be derived:}
    \begin{align}
        n_c &= \frac{q_n \rho_a}{m_\mathrm{n}}\\
        m_\mathrm{avg} &= \frac{q_c m_n}{q_n} \\
        \label{eq:radius}
        r &= \sqrt[3]{ \frac{3 m_\mathrm{avg}}{4 \pi \rho_c}}
    \end{align}
    \referee{The microphysical parameters of the cloud particle materials are listed in Table~\ref{tab:microparam}.}

    \referee{\texttt{Nimbus} uses a zero-flux (closed box) boundary condition for gas-phase and cloud particles at the top of the atmosphere. Equivalent to \texttt{Virga}, a constant gas-phase MMR below the cloud-forming layers ($S < 1$) is assumed, called deep MMR, which is a free parameter within \texttt{Nimbus}. The deep MMR accounts for both the diffusive replenishment from the interior and evaporating cloud particles below the cloud deck. Mass conservation between diffusive replenishment of gas-phase material and settling cloud particles is reached in the steady state.}
    
     
    \referee{A conceptual comparison between \texttt{Virga} and \texttt{Nimbus} is shown in Fig.~\ref{fig:vig_nimb_comp}. To compare \texttt{Nimbus} to other cloud models for exoplanet atmospheres, we present a summary of their assumptions in Table~\ref{tab:cloud_models}. It is important to note that Table~\ref{tab:cloud_models} does not represent an exhaustive list of exoplanet cloud models.}

    \subsection{The numerics behind \texttt{Nimbus}}
    \label{sec:app_nimbus}

    \begin{figure}
        \centering
        \includegraphics[width=\linewidth]{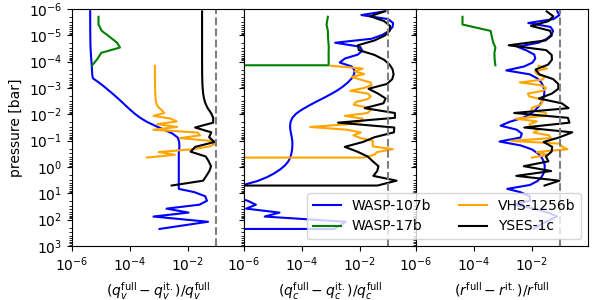}
        \caption{Relative differences between the gas-phase MMR $q_v$, cloud particle MMR $q_c$, and cloud particle radius $r$ between a full and iterative (it.) \texttt{Nimbus} run. The dashed line marks 10\% difference.}
        \label{fig:app_nimbus_test}
    \end{figure}
    
    \referee{\texttt{Nimbus} solves Eq.~\ref{eq:nimb_1} to \ref{eq:nimb_3} using a first-order advection and diffusion scheme. The numerical evaluation is done with the `LSODA' method of the solve\_ivp function from SciPy \citep{virtanen_scipy_2020} which is based on DLSODES \citep{hindmarsh_odepack_1983, radhakrishnan_description_1993}. It uses sparse matrix techniques to solve systems of stiff and non-stiff ODEs. To consider multiple cloud materials, \texttt{Nimbus} is run once for each material individually.} All \texttt{Nimbus} runs within this paper:
    \begin{itemize}
        \item use a relative tolerance of $r_\mathrm{tol} = 10^{-6}$,
        \item use an absolute tolerance of $a_\mathrm{tol} = 10^{-25}$,
        \item set all MMRs below $10^{-30}$ to 0,
        \item assume spherical particles, and
        \item assume a CCN radius of $r_\mathrm{n} = 10^{-3}$~$\mu$m.
    \end{itemize}
    \referee{The gas-phase chemistry affects cloud formation and vice versa. This creates feedback loops which lead to a stiff numerical problem. To prevent this, \texttt{Nimbus} includes an iterative method which uses the following steps:}
    \begin{enumerate}
        \setcounter{enumi}{-1}
        \item \referee{Use an initial $f_\mathrm{sed}$ value to calculate the cloud particle radii for each pressure layer $r(p)$.}
        \item \referee{Evaluate Eq.~\ref{eq:nimb_1}-\ref{eq:nimb_3} with a fixed value for $r(p)$ until a steady state is reached.}
        \item \referee{Recalculate $r(p)$  and smooth it using an 8th order polynomial to prevent numerical artifacts.}
        \item \referee{Continue with step 1, except if either:}
        \begin{enumerate}
            \item \referee{a given number of iterations is reached, or}
            \item \referee{the maximum relative difference for all MMRs is less than $10^{-3}$.}
        \end{enumerate}
    \end{enumerate}
    \referee{We use the iterative method in all \texttt{Nimbus} simulations. During the $\chi^2$ minimisation, a maximum of 10 iterations are performed. Most iterative \texttt{Nimbus} runs have runtimes between 10 to 20 seconds. This is slightly slower than typical runtimes of \texttt{Virga} ($\sim 1$~second) and \texttt{ExoLyn} \citep[$\sim 2$~seconds;][]{huang_exolyn_2024} but significantly faster than \texttt{CARMA} \citep[$\sim 10$~hours;][]{mang_microphysical_2024} and \texttt{DiffuDrift} \citep[$\sim$~days;][]{woitke_dust_2020}. To produce the final results, the `maximum relative difference' stopping criterion is used. These runs typically take between 10 to 50 seconds. We have found no significant difference in the $\chi^2$ values and the cloud structures between the two approaches. To confirm the accuracy of the iterative method, we compare it to full \texttt{Nimbus} runs for WASP-107b, WASP-17b, VHS-1256b, and YSES-1c. The relative differences can be seen in Fig.~\ref{fig:app_nimbus_test}.} \texttt{Nimbus} is available on GitHub and contributions are welcomed: \url{https://github.com/Kiefersv/Nimbus}.

    \subsection{Atmospheric mixing}
    \label{sec:Model_mixing}
    
    Mixing within atmospheres is caused by diffusion, convection, or advection: GCMs simulate the advective heat transport to predict the temperature structure of a planet \referee{\citep[e.g. for gas-giants and brown dwarfs:][]{showman_atmospheric_2002, showman_atmospheric_2009, carone_connecting_2014, mayne_unified_2014, komacek_vertical_2019, lee_simplified_2020, roman_clouds_2021, schneider_exploring_2022, steinrueck_photochemical_2023, tan_modelling_2024, teinturier_warm_2024};} Radiative-convective models simulate convective zones to achieve flux balance to determine the temperature structure \referee{\citep[e.g.][]{allard_synthetic_1996, marley_atmospheric_1996, fortney_comparative_2005, malik_helios_2017,  batalha_exoplanet_2019, marley_sonora_2021, windsor_radiative-convective_2023, mukherjee_picaso_2023, lacy_self-consistent_2023, morley_sonora_2024};} and chemistry models consider the molecular diffusion of gas-phase species to determine the gas-phase abundances throughout the atmosphere \citep[e.g.][]{hu_photochemistry_2012, moses_chemical_2014, tsai_vulcan_2017, rimmer_chemical_2016, baeyens_grid_2021, mukherjee_effects_2025}. One-dimensional models often summarise these processes within a single diffusion constant, $K_{zz}$.

    \referee{The $K_{zz}$ used in this paper are taken from different models depending on the planet. The values are shown in Fig.~\ref{fig:app_all_cloud_structures} and details about their derivation can be found in the respective original papers.}

    \subsection{Settling velocity}
    \label{sec:Model_vdr}
    
    \begin{figure}
       \centering
       \includegraphics[width=\linewidth]{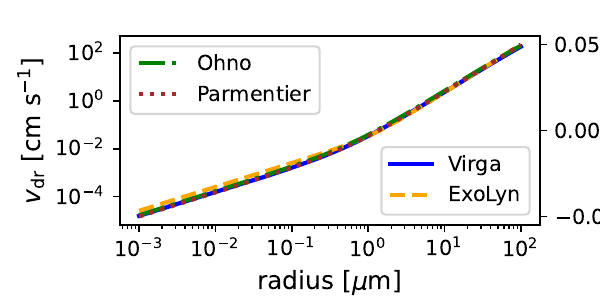}
       \caption{\referee{Comparison between settling velocity descriptions for a SiO particle at T = 1000 K, p = 1 bar, and log(g) = 3. The descriptions are taken from \citet{batalha_condensation_2026} (\texttt{Virga}), \citet{huang_exolyn_2024} (ExoLyn), \citet{ohno_clouds_2020}, and \citet{parmentier_3d_2013}.}}
       \label{fig:vsed_comp}
    \end{figure}
    
    The gravitational acceleration of cloud particles is counteracted by the frictional force of the surrounding gas. The balance between these two forces will determine the terminal settling velocity $v_\mathrm{dr}$ of the cloud particles. Since acceleration timescales are typically short \citep{woitke_dust_2003}, \referee{the cloud particles can be assumed to reach terminal velocity instantly.}
    
    The strength of the frictional force depends on the cloud particle and atmospheric properties. There are two regimes which can be distinguished using the Knudsen number:
    \begin{equation}
        Kn = \frac{l}{r}
    \end{equation}
    where $l$~[cm] is the mean free path. If the Knudsen number is large ($Kn \gg 1$), the drag force can be described by a free molecular flow. This is called the Epstein regime \citep{epstein_resistance_1924}. In the case of low Knudsen numbers ($Kn \ll 1$), there are two different regimes which can be distinguished by the Reynolds number:
    \begin{equation}
        Rn = \frac{2 r \rho_a v_\mathrm{dr}}{\mu_\mathrm{kin}}
    \end{equation}
    where $\mu_\mathrm{kin}$~[cm$^2$/s] is the kinetic viscosity of the gas. If the Reynolds number is low ($Rn \ll 1$), the drag force is best described by a laminar flow. This is called the Stokes regime \citep{stokes_effect_1851}. A high Reynolds number ($Rn \gg 1000$) indicates turbulent flow. The Cunningham slip factor provides a smooth transition between the low and high Reynolds number limit \citep{cunningham_velocity_1910}.

    \referee{For \texttt{Nimbus}, we use the $v_\mathrm{dr}$ interpolation scheme implemented in the GitHub version of ExoLyn \citet{huang_exolyn_2024}:}
    \begin{equation}
        v_\mathrm{dr} = \frac{g ~r ~\rho_c}{v_\mathrm{th}~\rho_a} \sqrt{1 + \left(\frac{4~r}{9~l} \right)^2}
    \end{equation}
    where $\rho_c$~[g/cm$^{3}$] is the solid density of the cloud particle, and $v_\mathrm{th} = \sqrt{8 R_g T/(\pi \mu)}$~[cm/s] is the thermal velocity. This interpolation scheme results in a smooth transition between the high and low Knudsen number regimes. The settling velocity of \texttt{Virga} is described in \citet{batalha_condensation_2026}. A comparison between \texttt{Virga} and \texttt{Nimbus} is shown in Fig.~\ref{fig:vsed_comp}.

    \subsection{Cloud particle size distribution}
    \label{sec:Model_PSD}
    
    \begin{figure}
        \centering
        \includegraphics[width=\linewidth]{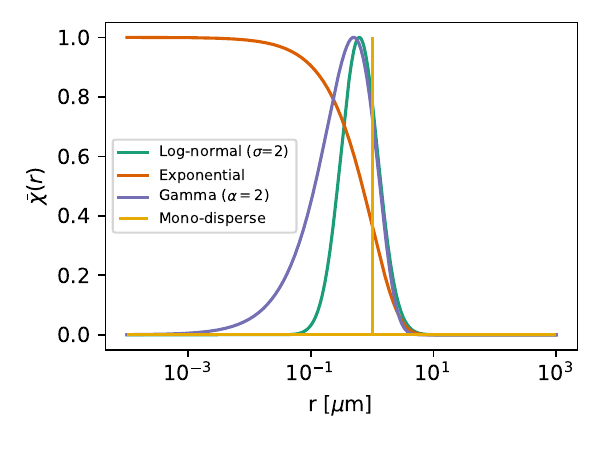}
        \caption{Re-normalised particle size distribution $\bar{\chi} = \chi/\mathrm{max(\chi)}$ as listed in Table~\ref{tab:size_dists} for $r_g = 10^{-4}$~cm. \texttt{Virga} assumes a log-normal distribution with $\sigma = 2$. \texttt{Nimbus} assumes mono-disperse cloud particles.}
        \label{fig:vis_size_dist}
    \end{figure}
    
    \begin{table}
        \centering
        \caption{Commonly assumed PSDs for clouds in exoplanet atmospheres. The variable $\alpha$ is a free, positive parameter. For $\alpha=1$ the gamma distribution is equal to the exponential distribution. The variable $\delta$ is the Dirac delta function.}
        \label{tab:size_dists}
        \begin{tabular}{l l l l}
            \hline\hline
             Distribution & PSD $\chi(r)$ [1/cm] & $\gamma_n$ \\  
             \hline 
            Log-normal & $\frac{1}{r \sigma \sqrt{2 \pi}} \exp \left( - \frac{\ln^2(r/r_g)}{2 \ln^2(\sigma)} \right)$ & $\exp \left(\frac{n^2 \ln^2(\sigma)}{2} \right)$ \\
            Exponential & $\frac{1}{r_g} \exp \left(- \frac{r}{r_g} \right)$ & n!\\
            Gamma & $\frac{r^{\alpha-1}}{\Gamma(\alpha) r_g^\alpha \alpha^{-\alpha}} \exp \left( - \frac{r}{r_g} \alpha \right)$ & $\frac{\Gamma(\alpha+n)}{\alpha^n \Gamma(\alpha)}$ \\
            Mono-disperse & $\delta(r-r_g)$ & 1 \\
            \hline
        \end{tabular}
    \end{table}

    \referee{Cloud particles in exoplanet atmospheres can have complex PSDs \citep[see e.g.][]{gao_microphysics_2018, powell_formation_2018, gao_aerosol_2020}. Most cloud models therefore use one of three approaches: cloud particle size bins, the moment method, or assuming a fixed size distribution.}
    
    \referee{Binning cloud particles into size bins and solving the cloud formation equations for each bin allows to account for size dependent growth effects and vertical differentiation of cloud particle sizes. Comprehensive size-bin models like CARMA \citep[e.g.,][]{gao_microphysics_2018, powell_formation_2018, gao_aerosols_2021, powell_two-dimensional_2024} have shown that size distributions can be multimodal and do not necessarily follow a simple distribution. However, the binning method is computationally expensive which limits its application for retrieval purposes.}

    \referee{The moment method \citep{gail_primary_1986, gail_dust_1988, dominik_dust_1993, helling_dust_2001, helling_dust_2006, lee_modelling_2023, lee_three-dimensional_2025} simplifies the cloud formation equations using the moments $L_j$ [cm$^J$/g]:}
    \begin{equation}
        L_j = \int_{r_\mathrm{min}}^\infty \chi(r)~r^j~dr
    \end{equation}
    \referee{where $r_\mathrm{min}$ [cm] is the minimum radius for a cloud particle. A detailed explanation of the moment method for exoplanet clouds can be found in \citet{helling_dust_2006}. This method is computationally fast and allows to consider heterogenous cloud particles. However, the physical properties of the cloud formation, like accretion and collisional growth rates, still require the assumption of a size distribution \citep{lee_beyond_2025, lee_beyonde_2025}.}

    \referee{The simplest approach is to assume a given PSD.} This has the advantage that complex integrals of cloud formation equations can be solved analytically. \referee{For example, the relationship between the cloud particle number density $n_\mathrm{c}$ and the total cloud particle mass mixing ratio $q_c$ can be simplified to:}
    \begin{equation}
         q_c = \frac{1}{\rho_a}\int_0^\infty \frac{4}{3} \pi r^3 \rho_c n_c  \chi(r) dr = \frac{\rho_c}{\rho_a} \frac{4}{3} \pi n_\mathrm{c} E_\chi[r^3]
    \end{equation}
    where $E_\chi[r^n]$ is the n-th moment of the function $\chi(r)$. The moments of the PSDs shown in Fig.~\ref{fig:vis_size_dist} are all well defined and can be expressed as:
    \begin{equation}
        \label{eq:size_dist_approx}
        E_\chi[r^n] = r_g^n\gamma_n
    \end{equation}
    where $r_g$~[cm] is the mean cloud particle radius and $\gamma_n$ is the PSD factor. \referee{Many models therefore assume one of the PSDs listed in Fig.~\ref{fig:vis_size_dist}.}
    
    \referee{For this paper, all \texttt{Virga} runs assume a log-normal PSD with $\sigma=2$. \texttt{Nimbus} uses a method similar to the moment scheme of \citet{lee_three-dimensional_2025}, \citet{ormel_arcis_2019}, and \citet{huang_exolyn_2024}, and assumes a mono-disperse PSD of homogenous particles.}

    \subsection{Nucleation rate}
    \label{sec:Model_nuc}
    Cloud formation in gas-giant planets starts with the formation of molecular clusters from the gas-phase. This process is called nucleation. The rate $J_s$ at which a given material $s$ nucleates can be calculated through kinetic nucleation theory but only if the thermodynamic properties of sufficiently larger clusters are known \citep{lee_dust_2015, boulangier_devloping_2019, boulangier_developing_2019, kohn_dust_2021, kiefer_fully_2024}. Unfortunately, only few species have enough data available to perform these calculations \citep{bromley_under_2016, sindel_revisiting_2022, gobrecht_bottom-up_2022, gobrecht_bottom-up_2023, lecoq-molinos_vanadium_2024}. Modified Classical Nucleation Theory (MCNT) provides an approximation of the nucleation rate in the absence of thermodynamic properties of larger clusters \citep{gail_physics_2013}. MCNT uses the surface tension $\sigma_\infty$ [erg/cm$^2$] to approximate the thermodynamic properties of the nucleating species:
    \begin{equation}
        \label{eq:nucrate}
        J = 4 \pi r^2 v_\mathrm{rel} ~Z(N_\star) ~n^2_1 ~ e^{\left((N_\star - 1) \ln S (T) - \Delta G(N_\star)/(k_BT) \right)}
    \end{equation}
    \referee{where $n_1$~[1/cm$^{3}$] is the number density of the nucleating species in the gas-phase, $v_\mathrm{rel}$~[cm/s] is the relative velocity, $Z$ is the Zeldovich factor, $N_\star$ the critical cluster size, $S(T)$ the supersaturation ratio, $k_b = 1.38\times10^{-16}$~erg/K is the Boltzmann constant, and $\Delta G$ [erg] the energy of formation.} The accuracy of this approximation depends on the nucleating species \citep{lee_dust_2015, bromley_under_2016, sindel_revisiting_2022, lecoq-molinos_vanadium_2024}. Examples on how to calculate the individual terms of Eq.~\ref{eq:nucrate} can be found in \citet{gao_sedimentation_2018}, and \citet{sindel_revisiting_2022}. \referee{Some materials nucleate more efficiently then others. For gas-giant exoplanets, SiO, TiO$_2$, and KCl are often considered the main nucleating species \citep{lee_dust_2015, bromley_under_2016, gao_microphysics_2018}. Other species, like Mg$_2$SiO$_4$, MgSiO$_3$, Fe, or ZnS, grow more efficiently onto these particles rather than nucleating themselves. This processes is called heterogenous nucleation \citep{yau_short_1996}.}
    
    \referee{
    For the planets studied in this work (Sect.~\ref{sec:obs_trans} and \ref{sec:obs_emis}), we scaled the nucleation rate with a factor $f_J$ to test its effect on the cloud structures. We found that orders of magnitude differences in the nucleation rate only lead to minor changes in the spectra (Sect.~\ref{sec:dis_cons_nucrate}). We therefore fix the nucleation rate for each cloud material within \texttt{Nimbus} to the homogenous MCNT rate of its species. While this potentially under predicts the nucleation rate of materials that could have formed through heterogenous nucleation (e.g. MgSiO$_3$, and Mg$_2$SiO$_4$), a comprehensive nucleation study is required to understand the impact of complex nucleation pathways on the cloud structure \citep{boulangier_devloping_2019, boulangier_developing_2019, kohn_dust_2021, gobrecht_bottom-up_2022, gobrecht_bottom-up_2023, kiefer_fully_2024}.
    }

    \subsection{Growth rate}
    \label{sec:Model_acc}

    Once CCNs are present in the gas-phase, other materials can start to accrete. In low density environments or for small cloud particles, the rate at which materials accrete $G_\mathrm{col}$~[1/cm$^{3}$/s] is limited by the number of collisions happening between gas-phase species and cloud particles:
    \begin{align}
        \label{eq:gcol}
        G_\mathrm{col} &= 4 \pi r^2~ s~v_\mathrm{c} ~n_1 ~n_{c}  ~\left(1 - \frac{p_\mathrm{vap}}{p_1} \right)
    \end{align}
    \referee{where $v_c = \sqrt{R_g T / (2 \pi m_c)}$~[cm/s] is the relative velocity of vapour molecules, $m_c$~[g/mol] is the molecular weight of the cloud forming gas-phase species, $p_1$ [bar] is the partial pressure of the growth species, and $p_\mathrm{vap}$ [bar] is the vapour pressure of the growth species.} In high density environments or for large cloud particles, the cloud forming materials are efficiently depleted near the cloud particle. The growth rate $G_\mathrm{dif}$~[1/cm$^{3}$/s] is therefore limited by the supply of new material through diffusion:
    \begin{align}
        \label{eq:gdif}
        G_\mathrm{dif} &= 4 \pi r ~D ~n_1 ~n_{c} ~\left(1 - \frac{p_\mathrm{vap}}{p_1} \right)
    \end{align}
    where $D$~[cm$^2$/s] is the gas phase diffusion constant. The derivation of $D$ can be found in \citet{jacobson_fundamentals_2005} \citep[see also][]{gao_microphysics_2018, ohno_microphysical_2018}. The variable $0 < s \leq 1$ in Eq.~\ref{eq:gcol} denotes the sticking coefficient. By default, we assume a sticking coefficient of 1 which is in line with previous studies \citep[see e.g.][]{lazzati_non-local_2008, bromley_under_2016, boulangier_devloping_2019, gao_aerosol_2020, kiefer_effect_2023}. However, we will show later, this assumption is not always accurate and lift it where necessary. 
    
    Some cloud models assume only one accretion regime \citep[e.g.][]{gao_microphysics_2018, ormel_arcis_2019, woitke_dust_2020, huang_exolyn_2024} where others interpolate between both \citep[see e.g.][]{ohno_microphysical_2018, helling_modelling_2013, lee_modelling_2023}. The interpolation function differs between models. \referee{For \texttt{Nimbus}, we use the tanh interpolation function for homogenous, mono-dispersed particles from \citet{lee_modelling_2023}:}
    \begin{align}
        f_x &= \frac{1}{2} \left( 1 - \tanh\left( 2 \log_{10}\left( \frac{G_\mathrm{dif}}{G_\mathrm{col}} \right) \right) \right) \\
        G &= G_\mathrm{dif} f_x + G_\mathrm{col} (1 - f_x)
    \end{align}

    Some cloud particle materials exist in the gas-phase and can condense\footnote{The phase transition from gas-phase to cloud particle is commonly referred to as condensation, even for solid cloud particles, where desublimation would be more accurate.} directly onto cloud particles. This is true, for example, for SiO[s] which can form through the following reaction:
    \begin{equation}
        \label{eq:chm_sio}
        \mathrm{SiO} \rightarrow \mathrm{SiO[s]}.
    \end{equation}
    Other cloud particles will form through more complex surface reactions \citep{gail_dust_1988, patzer_dust_1998, visscher_atmospheric_2010, helling_sparkling_2019, kiefer_fully_2024}. For example, SiO$_2$[s], MgSiO$_3$[s], and Mg$_2$SiO$_4$[s] preferentially form through \citep{helling_dust_2006, visscher_atmospheric_2010, grant_jwst-tst_2023}
    \begin{align}
        \label{eq:chm_sio2}
        \mathrm{SiO} + \mathrm{H_2O} &\rightarrow \mathrm{SiO[s]} + \mathrm{H_2} \\
        \label{eq:chm_mgsio3}
        \mathrm{Mg} + 2\mathrm{H_2O} + \mathrm{SiO} &\rightarrow \mathrm{MgSiO_3}[s] + 2\mathrm{H_2} \\
        \label{eq:chm_mg2sio4}
        2\mathrm{Mg} + 3\mathrm{H_2O} + \mathrm{SiO} &\rightarrow \mathrm{Mg_2SiO_4}[s] + 3\mathrm{H_2}
    \end{align}
    The rate of these surface reactions, and therefore the vapour pressure, depends on the metallicity of the gas-phase. The vapour pressures of SiO is taken from \citet{lee_modelling_2023}, MgSiO$_3$ and Mg$_2$SiO$_4$ are taken from \citet{visscher_atmospheric_2010}, and SiO$_2$ is taken from \citet{grant_jwst-tst_2023}.

    \subsection{The TwoPop approach}
    \label{sec:Model_twopop}

    \referee{In Sect~\ref{sec:obs_emis}, we will show that a singular cloud profile cannot explain the effect of clouds on the photosphere and the Si-O absorption feature simultaneously. To reproduce the observed spectra, we introduce the two-population (TwoPop) approach where we combine a base cloud deck (`base') and an extended cloud (`Ext.'). Both cloud structures are calculated independently, each from a normal \texttt{Virga} or \texttt{Nimbus} run. Each cloud structure has its own cloud particle MMR, $f_\mathrm{sed}$ (in case of \texttt{Virga}), and sticking coefficient (in case of \texttt{Nimbus}). For this paper, the extended cloud is always chosen to be made from MgSiO$_3$, and has a lower $f_\mathrm{sed}$ or sticking coefficient $s$ then the base cloud. In the case that both the base cloud and the extended cloud are made from MgSiO$_3$, the total $\mathrm{MMR_\mathrm{MgSiO_3}}$ is the sum of both clouds structures.}

    \subsection{Transmission and thermal emission spectra}

    All model transmission and thermal emission spectra presented here are produced with \texttt{PICASO} \citep{batalha_exoplanet_2019, mukherjee_picaso_2023}. The following gas-phase opacities have been used: 
    H$_2$O \citep{polyansky_exomol_2018}, 
    CO$_2$ \citep{huang_reliable_2014}, 
    CH$_4$ \citep{yurchenko_vibrational_2013, yurchenko_exomol_2014}, 
    NH$_3$ \citep{yurchenko_variationally_2011, wilzewski_h2_2016}, 
    N$_2$ \citep{rothman_hitran2012_2013}, 
    CO \citep{rothman_hitemp_2010, gordon_hitran2016_2017, li_accounting_2015}, 
    TiO \citep{mckemmish_exomol_2019, gharib-nezhad_exoplines_2021}, 
    VO \citep{mckemmish_exomol_2016, gharib-nezhad_exoplines_2021}, 
    and 
    FeH \citep{dulick_line_2003, hargreaves_high-resolution_2010}. All cloud particle opacities are calculated using the opacity calculation routines of \texttt{Virga}. This includes opacities from \texttt{Nimbus} cloud structures.

    \referee{
    To find the best fit values of our models for transmission spectra, we use reduced $\chi^2$ minimisation. For thermal emission spectra, we first bin the flux values into 100 logarithmically spaced bins. The error of each bin is conservatively estimated to be either the mean error or the standard deviation of the binned values, which ever is higher. The binned flux of both data and model is normalised:
    }
    \begin{equation}
        F_N = \frac{F}{\max(F)}
    \end{equation}
    \referee{
    where $F$~[erg/(cm$^{2}$~s~cm)] is the flux, and $F_N$ is the normalised flux. Because $F_N$ can span two orders of magnitude (see Sect.~\ref{sec:obs_emis}), we use a log-$\chi^2$-minimisation for the thermal emission spectra with the likelihood function defined as:
    }
    \begin{equation}
        \label{eq:log_chi_2}
        \chi^2_\mathrm{log} = \frac{1}{N}\sum_i\frac{\log_{10}(F_N^\mathrm{data}) - \log_{10}(F_N^\mathrm{model})}{\log_{10}(F_N^\mathrm{error})}
    \end{equation}
    \referee{
    where $N$ is the number of data points. For completeness, we also calculate the reduced $\chi^2$ values for each model. Since we are especially interested in the quality of fit to the Si-O feature, we evaluate the $\chi^2_{5.2}$ value where only data points beyond 5.2~$\mu$m are taken into account.
    }
    
    \referee{
    To estimate the sensitivity of the model spectra to each fitting parameter, we vary each parameter individually until the goodness-of-fit increases by 20\% relative to the best fit value. The relative offsets are noted as super- and subscripts for increases and decreases, respectively.
    }

\section{Transmission spectroscopy}
\label{sec:obs_trans}

    \begin{table*}
        \caption{Planetary and stellar parameters used in this work.}
        \label{tab:planet_param}
        \begin{tabular}{l l l l l l l l l}
            \hline\hline
            Planet & $M_p$ [$M_\mathrm{Jup}$] & $R_p$ [$R_\mathrm{Jup}$] & log($g_p$) [dex] & $R_S$ [$R_\odot$] & $T_S$ [K] & [Fe/H]$_S$ & log($g_S$) [dex] & Sources \\  
            \hline 
            WASP-107~b & 0.12 & 0.94 & 2.49 & 0.67  & 4425 & +0.02 & 4.633 & 1 \\
            WASP-17~b  & 0.63 & 1.83 & 2.67 & 1.572 & 6550 & -0.19 & 4.149 & 2 \\
            VHS-1256~b & 2.17 & 1.34 & 3.5  & -     & -    & -     & -     & 3 \\
            YSES-1~c   & 7.2  & 1.1  & 3.3  & -     & -    & -     & -     & 4 \\
            \hline
        \end{tabular}
        \tablecomments{(1) Planet data: \citet{anderson_discoveries_2017}. Star data: \citet{piaulet_wasp-107bs_2021}. (2) Planet data \citet{grant_jwst-tst_2023}. Star data: \citet{bonomo_gaps_2017}. Original values from \citet{anderson_wasp-17b_2009}. (3) Planet data: full Sonora model of \citet{petrus_jwst_2024}. (4) Mass of YSES-1~c taken from \citet{wood_tess_2023}. Radius of YSES-1~c taken from \citet{bohn_young_2020}. Surface gravity of YSES-1~c taken from \citet{zhang_eso_2024}.}
    \end{table*}
    
    \begin{figure}
        \centering
        \includegraphics[width=\linewidth]{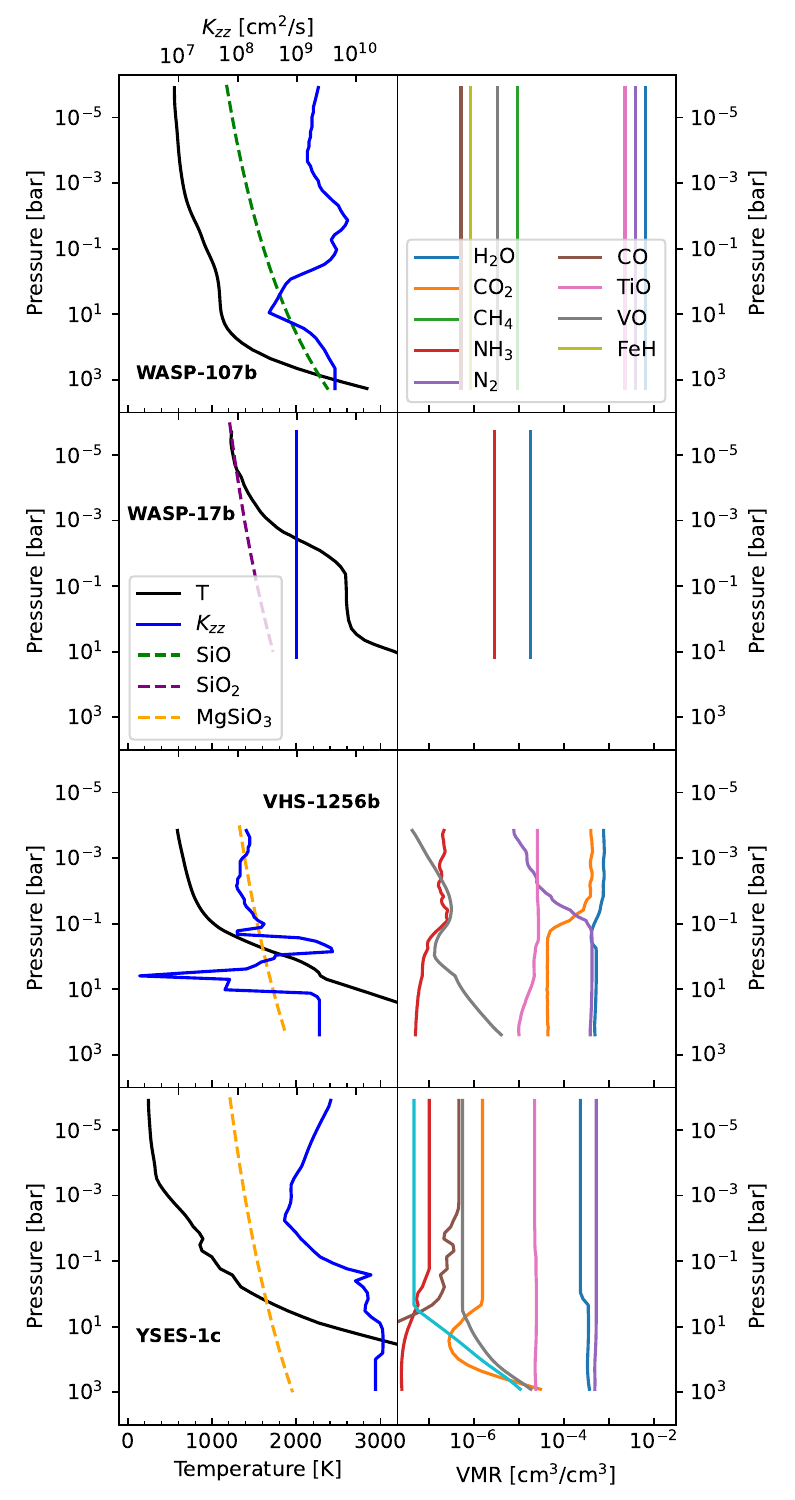}
        \caption{Atmospheric structure of WASP-107b ($T$-$p$ and $K_{zz}$: \citet{kreidberg_water_2018}; gas-phase abundances \citet{dyrek_so2_2023}), WASP-17b \citep{grant_jwst-tst_2023}, VHS-1256b \citep{zhou_roaring_2022}, and YSES-1c \citep{hoch_silicate_2025}. Also shown are the condensation curves of SiO, SiO$_2$, and MgSiO$_3$.}
        \label{fig:app_all_cloud_structures}
    \end{figure}

    In this section, we constrain the cloud structures of WASP-107~b \citep{anderson_discoveries_2017, dyrek_so2_2023, welbanks_high_2024} and WASP-17~b \citep{anderson_wasp-17b_2009, grant_jwst-tst_2023}. We use \texttt{Virga} to gain insights into the particle sizes and cloud particle MMR and \texttt{Nimbus} to gain insights into the accretion and nucleation rates. The results shown in this section are discussed in Sect.~\ref{sec:Discussion}.

    \subsection{WASP-107~b}
    \label{sec:WASP-107b}
    
    \begin{figure*}
        \centering
        \includegraphics[width=\linewidth]{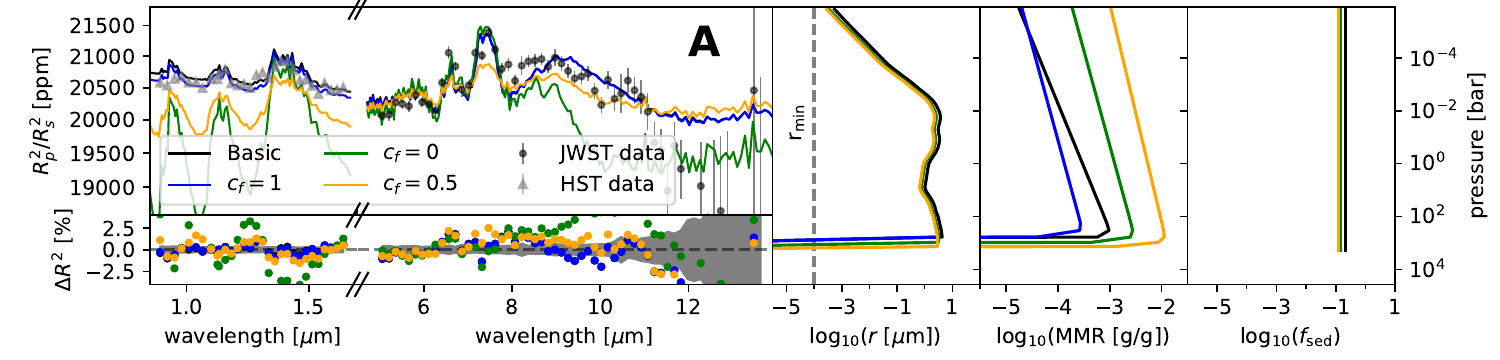}
        \includegraphics[width=\linewidth]{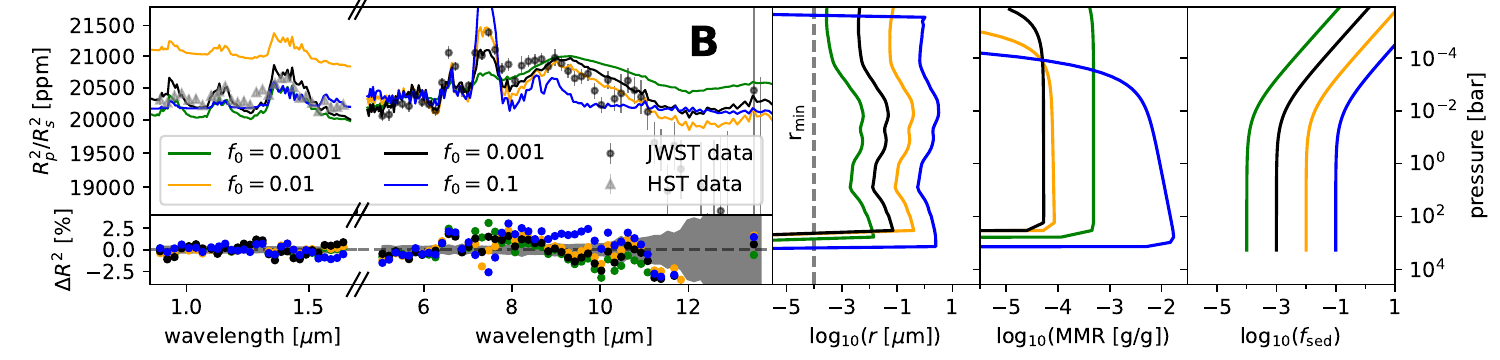}
        \includegraphics[width=\linewidth]{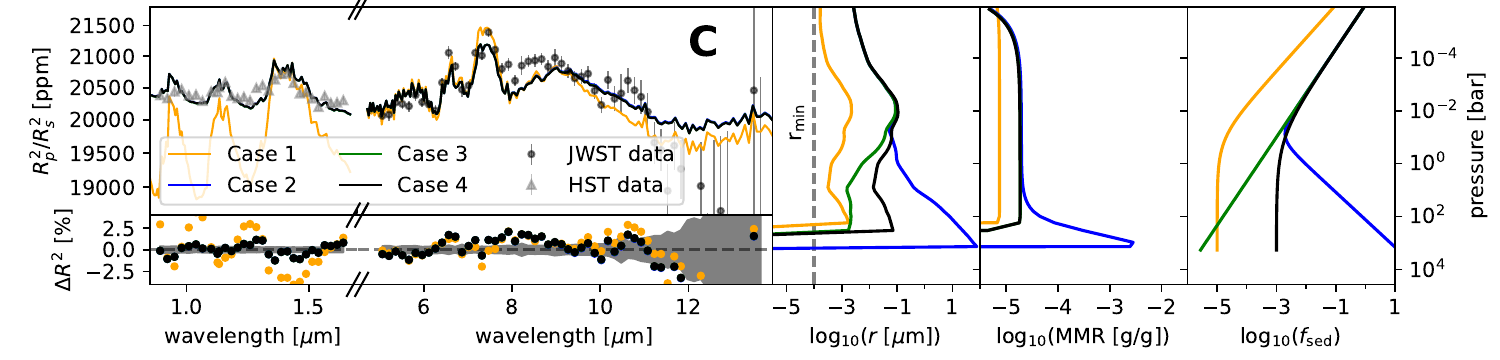}
        \includegraphics[width=\linewidth]{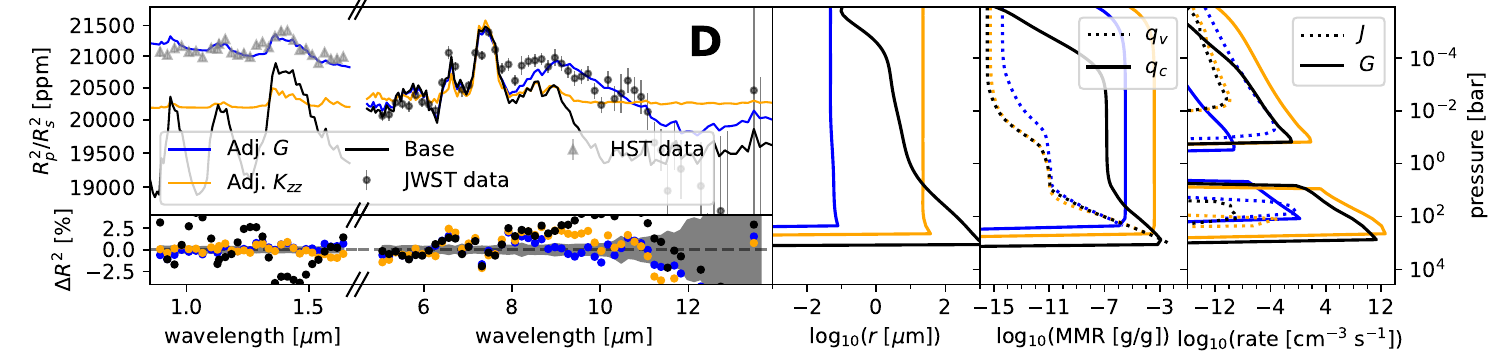}
        \caption{Transmission spectra and cloud structures of WASP-107~b. The cloud models of panel A, B, and C are calculated with \texttt{Virga}; the cloud models of panel D are calculated with \texttt{Nimbus}. The JWST and HST data are taken from \citep{welbanks_high_2024}.}
        \label{fig:wasp107b}
    \end{figure*}
    
    \begin{figure}
        \centering
        \includegraphics[width=1\linewidth]{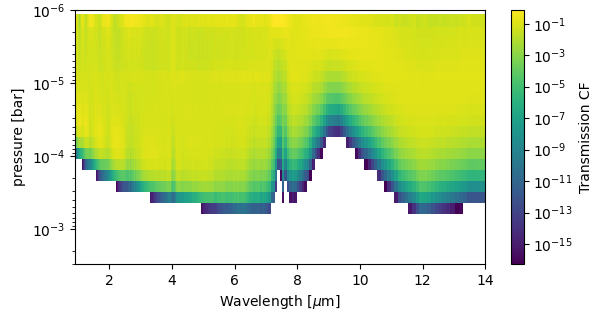}
        \caption{Contribution function for the transmission spectrum of WASP-107~b from panel A of Fig.~\ref{fig:wasp107b} for a cloud fraction of $c_f = 1$ and a MMR$_\mathrm{SiO} = 4.27 \cdot 10^{-4}$.}
        \label{fig:wasp107b_cf}
    \end{figure}
        
    \begin{table*}
        \centering
        \caption{Parameters for the cloud structure simulations of WASP-107~b with \texttt{Virga} and \texttt{Nimbus}. Boldfaced values are derived through $\chi^2$-minimisation. The `*' marker denotes entries with multiple values due to multiple cloud profiles being used. \referee{The following abbreviation is used: 1.23e-4 = $1.23 \cdot 10^{-4}$.}}
        \label{tab:wasp107b_tests}
        \begin{tabular}{l l l l l l l l l l l l l}
            \hline\hline
              & case            & $c_f$ & $f_0$ & $\alpha$ & $\beta$ & MMR$_\mathrm{SiO}$ & $\gamma_\mathrm{H_2O}$ & $\Delta_\mathrm{HST}$ [ppm] & $\chi^2$ \\  
            \hline 
            A & Basic           & 1     & \textbf{0.21}$^{+6.7\textnormal{e}-2}_{-5.4\textnormal{e}-2}$        & 0   & 0    & 1.14e-3                              & \textbf{7.86} & \textbf{+27}  & 3.83 \\
              & $c_f = 1$       & 1     & \textbf{0.14}$^{+6.5\textnormal{e}-2}_{-4.9\textnormal{e}-2}$        & 0   & 0    & \textbf{4.27e-4}$^{+6.2\textnormal{e}-4}_{-2.2\textnormal{e}-4}$ & \textbf{8.87} & \textbf{-56}  & 3.80 \\
              & $c_f = 0.5$     & 0.5   & \textbf{0.12}$^{+1.6\textnormal{e}-1}_{-1.0\textnormal{e}-2}$        & 0   & 0    & \textbf{1.26e-2}$^{+2.9\textnormal{e}-1}_{-1.2\textnormal{e}-2}$ & \textbf{19.8} & \textbf{-503} & 6.82 \\
              & $c_f = 0$       & 0     & -                                          & -   & -    & -                                                 & \textbf{10.0} & \textbf{-958} & 35.1  \\
            \hline
            B & $f_0 = 10^{-4}$ & 1     & $10^{-4}$            & 0.8 & 0    & \textbf{4.89e-4}$^{+1.2\textnormal{e}-2}_{-4.5\textnormal{e}-4}$ & \textbf{12.2} & \textbf{-490} & 6.19 \\
              & $f_0 = 10^{-3}$ & 1     & $10^{-3}$            & 0.8 & 0    & \textbf{5.36e-5}$^{+2.2\textnormal{e}-4}_{-3.9\textnormal{e}-5}$ & \textbf{8.25} & \textbf{-324} & 4.57 \\
              & $f_0 = 10^{-2}$ & 1     & $10^{-2}$            & 0.8 & 0    & \textbf{8.65e-5}$^{+5.4\textnormal{e}-4}_{-6.7\textnormal{e}-5}$ & \textbf{4.15} & \textbf{+388} & 3.68 \\
              & $f_0 = 10^{-1}$ & 1     & $10^{-1}$            & 0.8 & 0    & 1.88e-2                            & \textbf{0.27} & \textbf{-423} & 4.57 \\
            \hline
            C & Case 1          & 1     & $10^{-5}$            & 0.8 & 0    & $7.50 \cdot 10^{-6}$          & 10            & \textbf{-901} & 23.2 \\
              & Case 2          & 1     & $10^{-3}$            & 0.6 & 0.96 & $1.88 \cdot 10^{-2}$          & 14.1          & \textbf{-257} & 4.32 \\
              & Case 3          & 1     & $5 \cdot 10^{-4}$    & 0.6 & -0.6 & $1.88 \cdot 10^{-5}$          & 14.1          & \textbf{-260} & 4.36 \\
              & Case 4          & 1     & $10^{-4}$            & 0.6 & 0    & $1.88 \cdot 10^{-5}$          & 14.1          & \textbf{-258} & 4.37 \\
            \hline\hline
              & case            & $c_f$ & $s$ & $k_f$ & - & MMR$_\mathrm{SiO}$ & $\gamma_\mathrm{H_2O}$ & $\Delta_\mathrm{HST}$ [ppm] & $\chi^2$ \\
            \hline
            D & Base            & 1     & -                            &-&- & $1.00 \cdot 10^{-2}$          & 10            & \textbf{-867} & 27.2\\  
              & Adj. $G$        & 1     & \textbf{1.1e-7}$^{+7.2\textnormal{e}-6}_{-6.8\textnormal{e}-8}$ &-&- & \textbf{3.21e-6}$^{+4.0\textnormal{e}-6}_{-1.7\textnormal{e}-6}$ & \textbf{9.33} & \textbf{+413} & 3.31\\
              & Adj. $K_{zz}$   & 1     & - & \textbf{676}$^{+2.2\textnormal{e}+3}_{-4.3\textnormal{e}+2}$             &- & \textbf{4.00e-4}$^{+5.9\textnormal{e}-3}_{-3.8\textnormal{e}-4}$ & \textbf{0.22} & \textbf{-435} & 6.08\\
              & Combined        & 1     & * & *                        &- & *                             & 0.5           & \textbf{-7}  & 3.94\\
            \hline
        \end{tabular}
    \end{table*}

    WASP-107~b is a puffy warm Saturn (see Table~\ref{tab:planet_param}) that was previously observed with HST \citep{spake_helium_2018, kreidberg_water_2018} and chosen as one of the first targets to be observed with JWST \citep{dyrek_so2_2023, welbanks_high_2024, sing_warm_2024, murphy_evidence_2024}. The observational data\footnote{Available in MAST: \dataset[10.17909/8ffy-gs35]{https://doi.org/10.17909/8ffy-gs35}.} of this section is taken from \citet{welbanks_high_2024}. These observations show clear signs of clouds through muted molecular features and an Si-O absorption around 10~$\mu$m. Because \citet{dyrek_so2_2023} found that the observations are best explained with SiO clouds, we assume SiO to be the only cloud material for our study. The $T$-$p$ and $K_{zz}$ structure of WASP-107b were taken from \citet{kreidberg_water_2018} and can be seen in Fig.~\ref{fig:app_all_cloud_structures}. The gas-phase abundances were taken from the best fit model of \citet{dyrek_so2_2023}. Only the H$_2$O abundance is scaled by a factor $\gamma_\mathrm{H_2O}$ to fit the spectra. The additional planetary and stellar parameters used for the simulation are listed in Table~\ref{tab:planet_param}.

    \subsubsection{Constant \texorpdfstring{$f_\mathrm{sed}$}{f	extunderscore sed}}
    \label{sec:WASP-107b_cf}
    
    A common assumption made to model clouds with \texttt{Virga} is a constant $f_\mathrm{sed}$ throughout the atmosphere ($f_\mathrm{sed}(p) \equiv f_0$) and a fixed mass mass mixing ratio of the cloud forming material MMR$_\mathrm{SiO}$ [g/g] as listed in Eq.~\ref{eq:mmr_sio}. In addition to this `Basic' cloud structure, we use \texttt{Virga} with MMR$_\mathrm{SiO}$ [g/g] as a free parameter to simulate a fully cloudy planet with a cloud fraction $c_f$ of 1, a planet where only one terminator is cloudy ($c_f = 0.5$), and a cloudless planet ($c_f = 0$). The transmission spectrum, the cloud structures and the relative differences $\Delta R^2$ between the observations and the model are shown in panel A of Fig.~\ref{fig:wasp107b}. An offset $\Delta_\mathrm{HST}$ was applied to the HST data to achieve the best fit. The cloud parameters, HST offsets, and reduced $\chi^2$ values are listed in Table~\ref{tab:wasp107b_tests}. 
    
    The `Basic' \texttt{Virga} model has a 2.7 times higher MMR$_\mathrm{SiO}$ than when the MMR$_\mathrm{SiO}$ is fitted. This finding implies that either there is less SiO than our models predict (due to chemistry or lower Si abundance) or less SiO than expected is condensing into a cloud. All three cloudy models have a SiO feature around 10~$\mu$m. However, when only half the planet is covered the muting of molecular lines in the HST wavelengths is insufficient. This confirms model expectations that WASP-107~b is likely to have global cloud coverage. In all cases, $f_\mathrm{sed}$ is smaller than 1, resulting in extended cloud structures with small particles ($r < 10~\mu$m). Both of these are necessary to produce the SiO feature. 
    
    Fig.~\ref{fig:wasp107b_cf} shows the contribution function of the $c_f = 1$ case. Here, we see that the Si-O bond feature is produced at high altitudes ($p < 10^{-4}$~bar). At wavelengths between 0.8 to 5~$\mu$m, the cloud particles mute the molecular water features and cause a characteristic slope through Rayleigh scattering \citep{pont_detection_2008, sing_hubble_2011, pont_prevalence_2013, sing_hst_2015}. There is no contribution to the transmission spectrum from deeper altitudes ($p > 10^{-3}$~bar). The physical properties of cloud particles near the base of the cloud (p $\approx 10^3$) are therefore not directly observable.

    \subsubsection{Cloud particle sizes}
    
    Spectral features of cloud particles can only occur if the size of the particle is smaller than the wavelength of the spectral features \citep{wakeford_transmission_2015, hoch_silicate_2025}. The occurrence of the Si-O bond features in WASP-107~b thus limits the cloud particle sizes in the observable pressure range to $r < 10$~$\mu$m. The lower limit for the cloud particle sizes is given by the size of molecules ($r > 10^{-4}$~$\mu$m). It is unclear how many molecules are needed before a cluster starts to behave like a solid particle. Most studies estimate the number to be between 100 to 1000 \citep[see e.g.][]{kiefer_fully_2024, ormel_arcis_2019, huang_exolyn_2024}. The lower limit can therefore be conservatively estimated to be $r > 10^{-3}$~$\mu$m.
    
    To test if the sizes can be further constrained, we perform four \texttt{Virga} simulations with varying particle sizes. We use the following parametrisation of $f_\mathrm{sed}$ to achieve roughly constant cloud particle sizes in the upper atmosphere:
    \begin{equation}
        \label{eq:fsed_param}
         f_\mathrm{sed}(p) = f_0 \left(\left(\frac{p_0}{p} \right)^{\alpha} + \left(\frac{p}{p_0} \right)^{\beta} \right)
    \end{equation}
    where $\alpha$ and $\beta$ are hyperparameters, and $p_0$~[bar] is the reference pressure. For this section, we chose $\alpha = 0.8$, $\beta = 0$ and $p_0 = 0.01$~bar. This parametrisation leads to $f_\mathrm{sed}$ increasing with altitude which is in agreement with the results of \citet{rooney_new_2022}. The results are shown in Table~\ref{tab:wasp107b_tests} and panel B of Fig.~\ref{fig:wasp107b}.
    
    For $f_\mathrm{0} = 10^{-4}$, the cloud particles approach their lower size limit. For $f_\mathrm{0} = 10^{-1}$, the silicate feature starts to vanish as the cloud particles sizes approach 10~$\mu$m. All values between these two limits produce a reasonable fit to the JWST MIRI transmission spectrum of WASP-107~b. We were therefore not able to constrain the particle size further using \texttt{Virga}.

    \subsubsection{Deep cloud particle MMR}
    \label{sec:WASP-107b_mmr}
    
    In one-dimensional models, cloud particles can only reach higher altitudes through diffusive mixing described by $K_{zz}$. The cloud particle MMR can thus only decrease with increasing altitude. This allows us to determine the lower limit of the deep MMR by calculating the minimum MMR in the upper atmosphere needed to produce the observed transmission spectrum. 
    
    The opacity of cloud particles increases with the radius squared whereas the mass increases with the radius cubed. The highest opacity with the lowest mass is therefore achieved with the smallest particles. For WASP-107~b we find that a MMR$_\mathrm{SiO} = 1.02 \times 10^{-5}$ can still produce the silicate feature if cloud particles approach the lower size limit (See case 1 in panel C of Fig.~\ref{fig:wasp107b}). However, these particles alone do not sufficiently mute the water features in the HST wavelength range.
    
    The contribution function from Fig.~\ref{fig:wasp107b_cf} shows that cloud particles at pressures higher than 10$^{-3}$~bar do not contribute to the transmission spectrum. We attempt to constrain the radius and MMR$_\mathrm{SiO}$ below 10$^{-3}$~bar using three \texttt{Virga} runs. For each run we adjusted $f_\mathrm{sed}$ to achieve the same MMR$_\mathrm{SiO}$ and cloud particle size in the upper atmosphere ($p < 10^{-3}$~bar) but a different cloud base structure. The results are shown in Table~\ref{tab:wasp107b_tests} and panel C of Fig.~\ref{fig:wasp107b}. Using different $f_\mathrm{sed}$ we are able to vary the cloud particle radius at the cloud base by over four orders of magnitude and the mass mixing ratio by three orders of magnitude without inducing any changes in the transmission spectrum. This shows the need for microphysical cloud models which can help to achieve more rigorous constraints.

    \subsubsection{Accretion and mixing timescales}
    \label{sec:WASP-107b_nimbus}

    We use \texttt{Nimbus} to model the cloud structures of WASP-107~b. However, a standard \texttt{Nimbus} model run shows no Si-O absorption nor muting of the water features because not enough cloud material reaches the upper atmosphere (See `Base' in panel D of Fig.~\ref{fig:wasp107b}). Within the limitations of a one-dimensional models, there are two ways to correct this: decrease the growth rate or increase the mixing. We test both solutions. The results are shown in Table~\ref{tab:wasp107b_tests} and panel D of Fig.~\ref{fig:wasp107b}.

    The growth rate $G$ can be decreased by reducing the sticking coefficient $s$. \referee{To achieve a good fit to the data, the sticking coefficient must be smaller than $10^{-7}$.} While the resulting cloud structure reproduces the Si-O feature as well as the muting of the HST water lines, the sticking coefficient is smaller then expected from experiments ($s > 10^{-4}$) but within lower limits of theoretical predictions (See Section~\ref{sec:dis_twopop_s}).

    To study how stronger vertical mixing affects the cloud structure, we multiply the mixing rate $K_{zz}$ by a scaling factor $k_f$. A higher mixing rate brings more cloud material into the upper atmosphere \citep{samra_clouds_2022}. However, because accretion remains efficient, the particles become larger than 10~$\mu$m. These particles do not produce the Si–O feature and cannot explain the observations. The results of WASP-107~b are further discussed in comparison to the other planets in Section \ref{sec:Discussion}.

    \subsection{WASP-17~b}
    \label{sec:WASP-17b}
    
    \begin{figure*}
        \centering
        \includegraphics[width=\linewidth]{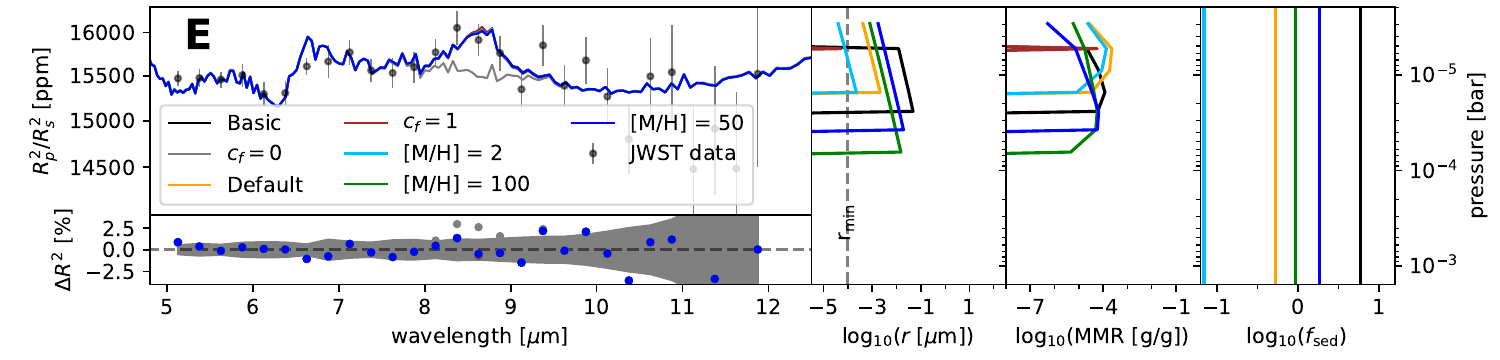}
        \includegraphics[width=\linewidth]{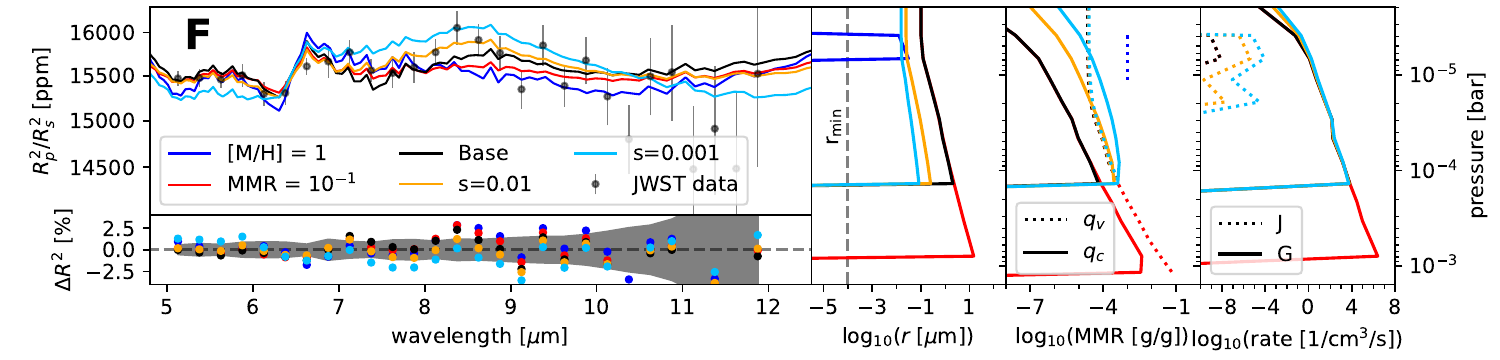}
        \caption{\referee{Transmission spectra and cloud structures of WASP-17~b. The cloud models of panel E are calculated with \texttt{Virga}; the cloud models of panel F are calculated with \texttt{Nimbus}.}}
        \label{fig:wasp17b}
    \end{figure*}
    
    \begin{table}
        \centering
        \caption{Parameters for the cloud structure simulations of WASP-17~b with \texttt{Virga} and \texttt{Nimbus}. Boldfaced values are derived through $\chi^2$-minimisation. \referee{The default metallicity is solar ([M/H] = 1) for \texttt{Virga} and [M/H] = 100 for \texttt{Nimbus}. All runs assume a cloud fraction of $c_f=0.5$ unless stated otherwise.} The following abbreviation is used: 1.23e-4 = $1.23 \cdot 10^{-4}$.}
        \label{tab:wasp17_tests}
        \begin{tabular}{l l l l l l l l l}
            \hline\hline
            & case            & $f_\mathrm{sed}$ & MMR$_\mathrm{SiO_2}$ & $\chi^2$ \\  
            \hline
            E & Default       & \textbf{0.53}$^{+4.2\textnormal{e}+0}_{-4.5\textnormal{e}-1}$ & \textbf{1.08e-3}$^{+1.1\textnormal{e}-4}_{-1.2\textnormal{e}-4}$ & 0.72 \\
              & Basic         & \textbf{6.00}$^{+1.3\textnormal{e}+1}_{-3.6\textnormal{e}+0}$ & 1.55e-3            & 0.72 \\
              & $c_f$=1       & \textbf{0.05}$^{+5.8\textnormal{e}-2}_{-1.1\textnormal{e}-2}$ & \textbf{8.21e-4}$^{+2.8\textnormal{e}-5}_{-2.8\textnormal{e}-5}$ & 0.73 \\
              & $c_f$=0      & -                             & -     & 1.13 \\
              & [M/H]=2       & \textbf{0.07}$^{+9.3\textnormal{e}-1}_{-2.3\textnormal{e}-2}$ & \textbf{3.84e-4}$^{+5.3\textnormal{e}-5}_{-5.5\textnormal{e}-5}$ & 0.72 \\
              & [M/H]=50      & \textbf{1.84}$^{+5.6\textnormal{e}+0}_{-1.4\textnormal{e}+0}$ & \textbf{1.26e-4}$^{+8.9\textnormal{e}-5}_{-6.6\textnormal{e}-5}$ & 0.72 \\
              & [M/H]=100     & \textbf{0.94}$^{+3.8\textnormal{e}+0}_{-8.3\textnormal{e}-1}$ & \textbf{7.98e-5}$^{+4.6\textnormal{e}-5}_{-4.3\textnormal{e}-5}$ & 0.72 \\
            \hline\hline
            & case & $s$ & MMR$_\mathrm{SiO_2}$ & $\chi^2$ \\ 
            \hline 
            F & Base          & 1             & $10^{-3}$            & 1.09 \\
              & [M/H]=50      & 1             & $10^{-3}$            & 1.08 \\
              & [M/H]=1       & 1             & $10^{-3}$            & 1.35 \\
              & MMR=$10^{-5}$ & 1             & $10^{-5}$            & 1.33 \\
              & MMR=$10^{-1}$ & 1             & $10^{-1}$            & 1.00 \\
              & $s$=0.01      & $10^{-2}$     & $10^{-3}$            & 0.72 \\
              & $s$=0.001     & $10^{-3}$     & $10^{-3}$            & 1.78 \\
            \hline
        \end{tabular}
    \end{table}

    WASP-17~b has an extended atmosphere favourable for transmission spectroscopy (see Table~\ref{tab:planet_param}). Observations with JWST MIRI\footnote{Available in MAST: \dataset[10.17909/19qv-5h62]{https://doi.org/10.17909/e61r-hk80}.} revealed spectral features of SiO$_2$ clouds in its atmosphere \citep{grant_jwst-tst_2023}. To model the cloud structure of WASP-17~b we therefore consider SiO$_2$ clouds and use the temperature structure from the \texttt{PICASO} forward model of \citet{grant_jwst-tst_2023} and the gas-phase abundances from their \texttt{petitRADTRANS} retrieval. The $T$-$p$, $K_{zz}$, and gas-phase abundances of WASP-17b were taken from \citet{grant_jwst-tst_2023} and can be seen in Fig.~\ref{fig:app_all_cloud_structures}. The additional planetary and stellar parameters used for the simulation are listed in Table~\ref{tab:planet_param}.
    
    Hot Jupiters are known to have strong equatorial wind jets which advect the hot dayside of the planet into the evening terminator, and the colder night side into the morning terminator \citep{baeyens_grid_2021, helling_exoplanet_2023}. Because of WASP-17~b's temperature structure, it is therefore likely that only the morning terminator has clouds.

    \subsubsection{Cloud fraction and metallicity}
    \label{sec:WASP-17b_vv}

    To investigate the cloud structure we perform six \texttt{Virga} runs. The `Default' simulation assumes a free MMR$_\mathrm{SiO_2}$, a cloud fraction of $c_f = 0.5$, and a metallicity of [M/H] = 1. Each other run tests the significance of a single assumption: a fixed MMR$_\mathrm{SiO_2}$ as listed in Eq.~\ref{eq:chm_sio2} (`Basic'), a fully cloudy planet $c_f = 1$ (`Full Cloud'), or a different metallicity [M/H] $\in \{ 1, 2, 50, 100 \}$. For comparison, we also produce a cloud free spectrum ($c_f = 0$). All results are listed in Table~\ref{tab:wasp107b_tests} and some results are shown in panel E of Fig.~\ref{fig:wasp17b}. The `Full Cloud' and `[M/H]=50' are omitted since they do not show significant differences to the `Default' and `[M/H]=100' case, respectively.
    
    All our cloud structures over fit the data, making it difficult to constrain $f_\mathrm{sed}$ and MMR$_\mathrm{SiO_2}$. In all cases, the cloud structure is limited to a small pressure range in the upper atmosphere. This is due to the high temperatures of WASP-17~b. Because SiO$_2$ forms through a surface reaction, higher gas-phase metallicities increase the vapour pressure of SiO$_2$ (See Sect.~\ref{sec:Model_acc}). Our results confirm that at higher metallicities clouds can form at lower altitudes leading to a more extended cloud structure. However, our best fit cloud particle MMR does not reflect the increase in metallicity and remains subsolar, even decreasing with increasing metallicity. However, it is difficult to derive any clear constraints as a large range of MMR$_\mathrm{SiO_2}$ achieve a $\chi^2 < 1$.

    \subsubsection{Insights from \texttt{Nimbus}}
    
    While we attempted to find the best fit \texttt{Nimbus} cloud structure via minimisation, increasing or decreasing the MMR$_\mathrm{SiO_2}$ by an order of magnitude did not change $\chi^2$ significantly ($\Delta \chi^2 < 0.025$). \referee{We therefore did not perform a $\chi^2$-minimisation but rather explored different MMR$_{\rm SiO_2}$, metallicity, and sticking coefficient values to study their impact on the transmission spectrum. The results are shown in Table~\ref{tab:wasp107b_tests} and panel F of Fig.~\ref{fig:wasp17b}.}
    
    \referee{We consider [M/H] = 100, MMR$_{\rm SiO_2} = 10^{-3}$, and $s = 1$ as the `Base' case of WASP-17~b. Deviating from these values, we find that:}
    \begin{enumerate}
        \item \referee{For solar metallicity ([M/H] = 1) or MMR$_{\rm SiO_2} = 10^{-5}$, the clouds do not impact the transmission spectrum. However, a metallicity of [M/H] = 50 already leads to a cloud structure that fits the data.}
        \item \referee{For MMR$_{\rm SiO_2} = 0.1$, the cloud particles grow too large to produce the Si-O feature. However, they still achieve a $\chi^2$ of 1.}
        \item \referee{A sticking coefficient of $s \approx 0.01$ leads to smaller cloud particles which better fit the Si-O bond feature. At lower values ($s \approx 0.001$), the Si-O feature becomes to broad to fit the data.}
    \end{enumerate}
    \referee{The \texttt{Nimbus} results reinforce our findings from Sect.~\ref{sec:WASP-17b_vv} that the observations of \citet{dyrek_so2_2023} are not sensitive to the details of the cloud structure.}

\section{Thermal emission spectroscopy}
\label{sec:obs_emis}

   The large wavelength coverage of JWST allows for detailed insights into the atmospheric structure of wide separation planets. In this section, we analyse VHS-1256~b \citep{gauza_discovery_2015, miles_jwst_2023} and YSES-1c \citep{bohn_two_2020, hoch_silicate_2025}, both have confirmed Si-O features around 10~$\mu$m. We use \texttt{Virga} to gain insights into the particle sizes and cloud particle MMR and \texttt{Nimbus} to gain insights into the accretion and nucleation rates. The results shown in this section are discussed in Sect.~\ref{sec:Discussion}.

    \subsection{VHS-1256~b}
    \label{sec:VHS-1256b}
    
    \begin{figure*}
    
        \centering
        \captionof{table}{Parameters for the cloud structure simulations of VHS-1256~b with \texttt{Virga} and \texttt{Nimbus}. Boldfaced values are derived through $\chi_\mathrm{log}^2$-minimisation. The following abbreviation is used: 1.23e-4 = $1.23 \cdot 10^{-4}$.}
        \label{tab:vhs1256b_tests}
        \begin{tabular}{l l l l l l l l l l l l l}
            \hline\hline
              & case    & base          & $f_\mathrm{base}$ & MMR$_\mathrm{base}$ & $f_\mathrm{0}$         & MMR$_\mathrm{MgSiO_3}$ & $\gamma_\mathrm{H_2O}$ & $\chi^2$ & $\chi^2_{5.2}$ & $\chi^2_\mathrm{log}$\\  
            \hline 
            G & Basic   & MgSiO$_3$     & \textbf{2.62}$^{+1.5\textnormal{e}-1}_{-1.4\textnormal{e}-1}$     & 2.59e-3             & -                      & -                      & \textbf{0.24}          & 4.02 & 4.67 & 1.3e-3 \\
              & Fit MMR & MgSiO$_3$     & \textbf{2.15}$^{+6.4\textnormal{e}-1}_{-6.9\textnormal{e}-2}$     & \textbf{1.61e-3}$^{+1.3\textnormal{e}-4}_{-9.2\textnormal{e}-5}$    & -                      & -                      & \textbf{0.29}          & 3.92 & 6.58 & 8.4e-4 \\
              & TwoPop  & MgSiO$_3$     & \textbf{2.00}$^{+4.4\textnormal{e}-2}_{-3.8\textnormal{e}-2}$     & \textbf{1.40e-3}$^{+5.9\textnormal{e}-5}_{-5.3\textnormal{e}-5}$    & \textbf{1.34e-4}$^{+7.8\textnormal{e}-5}_{-1.2\textnormal{e}-4}$       & \textbf{2.14e-5}$^{+1.5\textnormal{e}-5}_{-1.1\textnormal{e}-5}$       & \textbf{0.31}          & 1.28 & 0.29 & 3.8e-4 \\
            \hline
            H & -       & Mg$_2$SiO$_4$ & \textbf{2.36}$^{+5.4\textnormal{e}-2}_{-5.2\textnormal{e}-2}$     & \textbf{2.89e-3}$^{+3.3\textnormal{e}-4}_{-2.9\textnormal{e}-4}$    & \textbf{2.53e-5}$^{+1.4\textnormal{e}-4}_{-2.2\textnormal{e}-5}$       & \textbf{2.46e-5}$^{+1.9\textnormal{e}-5}_{-1.2\textnormal{e}-5}$       & \textbf{0.41}          & 0.97 & 0.33 & 3.2e-4 \\
              & -       & TiO$_2$       & \textbf{0.47}$^{+3.7\textnormal{e}-3}_{-3.7\textnormal{e}-3}$     & \textbf{6.20e-5}$^{+9.3\textnormal{e}-7}_{-9.2\textnormal{e}-7}$    & \textbf{1.80e-4}$^{+1.4\textnormal{e}-5}_{-1.5\textnormal{e}-5}$       & \textbf{4.27e-5}$^{+5.4\textnormal{e}-6}_{-5.4\textnormal{e}-6}$       & \textbf{0.21}          & 1.54 & 0.25 & 1.3e-4 \\
              & -       & Fe            & \textbf{1.31}$^{+2.4\textnormal{e}-2}_{-2.3\textnormal{e}-2}$     & \textbf{1.11e-3}$^{+3.0\textnormal{e}-5}_{-2.8\textnormal{e}-5}$    & \textbf{1.90e-5}$^{+1.3\textnormal{e}-4}_{-1.6\textnormal{e}-5}$       & \textbf{2.43e-5}$^{+2.4\textnormal{e}-5}_{-1.4\textnormal{e}-5}$       & \textbf{0.29}          & 1.14 & 0.42 & 3.8e-4 \\
              & -       & SiO           & \textbf{1.29$^{+2.4\textnormal{e}-2}_{-2.3\textnormal{e}-2}$}     & \textbf{8.05e-4}$^{+1.2\textnormal{e}-5}_{-1.1\textnormal{e}-5}$    & \textbf{1.95e-4}$^{+4.7\textnormal{e}-4}_{-6.0\textnormal{e}-5}$       & \textbf{2.49e-5}$^{+9.5\textnormal{e}-6}_{-9.4\textnormal{e}-6}$       & \textbf{0.27}          & 0.92 & 0.27 & 3.5e-4 \\
              & -       & SiO$_2$       & \textbf{9.07}$^{+1.1\textnormal{e}+0}_{-9.6\textnormal{e}-1}$     & \textbf{4.10e-1}$^{+7.6\textnormal{e}-2}_{-5.4\textnormal{e}-2}$    & \textbf{2.03e-2}$^{+3.7\textnormal{e}-3}_{-3.2\textnormal{e}-3}$       & \textbf{1.37e-5}$^{+1.7\textnormal{e}-6}_{-1.6\textnormal{e}-6}$       & \textbf{0.27}          & 4.27 & 3.23 & 6.8e-4 \\
              & -       & Al$_2$O$_3$   & \textbf{3.32}$^{+5.2\textnormal{e}-2}_{-5.0\textnormal{e}-2}$     & \textbf{2.77e-1}$^{+1.4\textnormal{e}-1}_{-1.1\textnormal{e}-1}$    & \textbf{2.02e-2}$^{+1.7\textnormal{e}-3}_{-1.6\textnormal{e}-3}$       & \textbf{1.48e-5}$^{+1.1\textnormal{e}-6}_{-1.0\textnormal{e}-6}$       & \textbf{0.17}          & 2.75 & 2.18 & 3.8e-4 \\
            \hline\hline
              & case    & base          & $s_\mathrm{base}$ & MMR$_\mathrm{base}$ & $s_{\mathrm{MgSiO_3}}$ & MMR$_\mathrm{MgSiO_3}$ & $\gamma_\mathrm{H_2O}$ & $\chi^2$ & $\chi^2_{5.2}$ & $\chi^2_\mathrm{log}$\\  
            \hline
            I & Base    & MgSiO$_3$     & \textbf{0.24}$^{+3,1\textnormal{e}-2}_{-2.4\textnormal{e}-2}$     & \textbf{1.85e-4}$^{+2,7\textnormal{e}-5}_{-2.4\textnormal{e}-5}$    & -                      & -                      & \textbf{0.60}          & 3.9 & 4.3 & 8.7e-4 \\
              & TwoPop  & MgSiO$_3$     & \textbf{0.22}$^{+2,0\textnormal{e}-2}_{-1.7\textnormal{e}-2}$     & \textbf{1.71e-4}$^{+2.1\textnormal{e}-5}_{-1.8\textnormal{e}-5}$    & \textbf{1.00e-10}$^{+2.1\textnormal{e}-4}_{-----}$      & \textbf{2.13e-6}$^{+3.4\textnormal{e}-6}_{-1.6\textnormal{e}-6}$       & \textbf{0.63}          & 3.0 & 0.36 & 6.1e-4 \\
            \hline
        \end{tabular}

        \vspace{5mm}
        
        \centering
        \includegraphics[width=\linewidth]{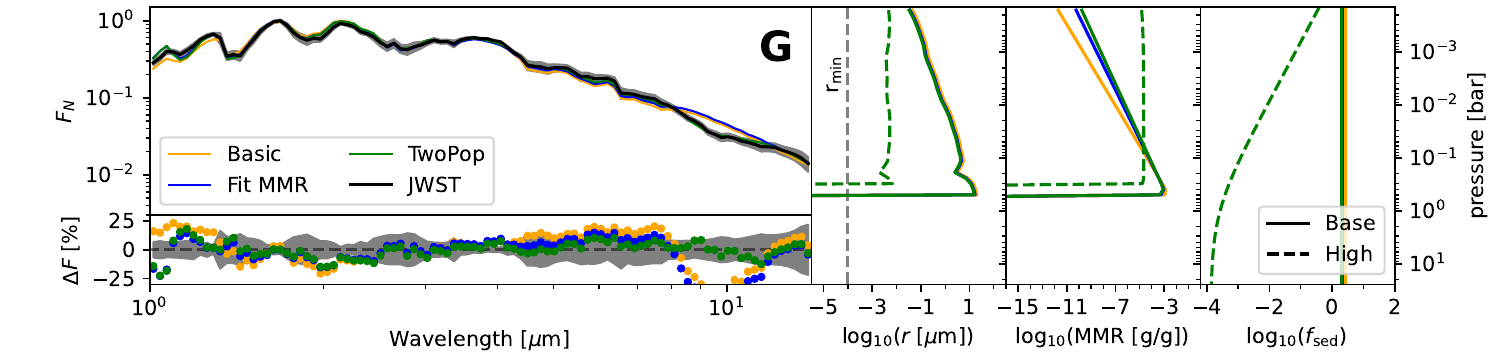}
        \includegraphics[width=\linewidth]{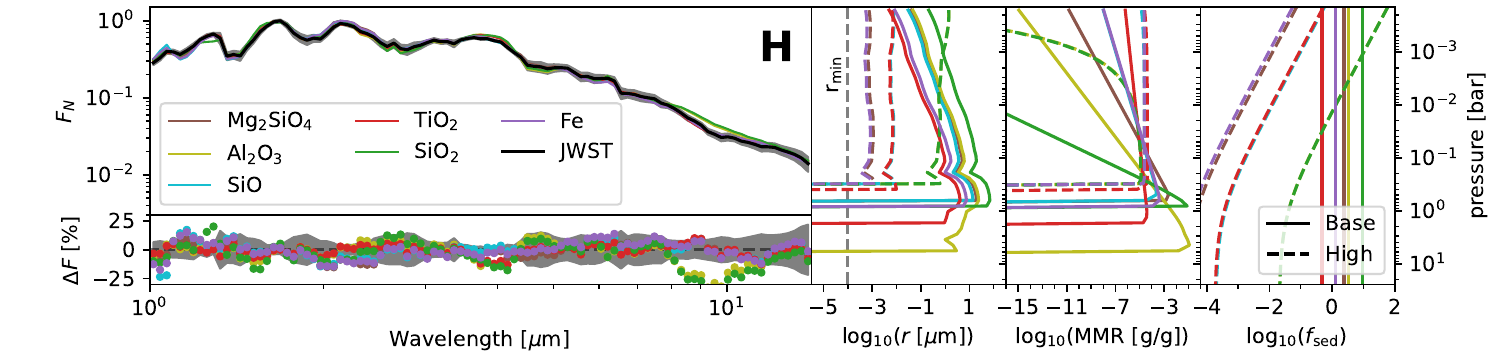}
        \includegraphics[width=\linewidth]{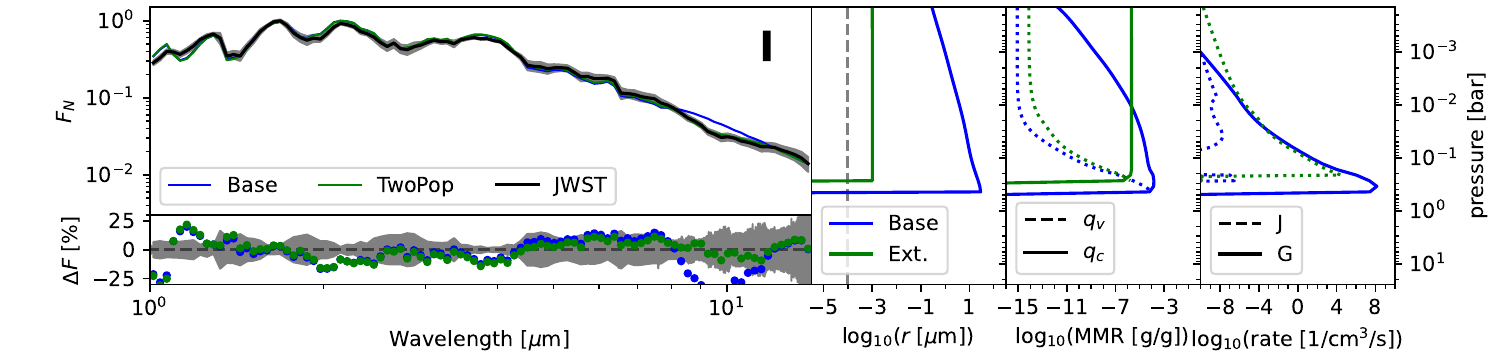}
        \captionof{figure}{Thermal emission spectra and cloud structures of VHS-1256~b. The cloud models of panel G and H are calculated with \texttt{Virga}; the cloud models of panel I are calculated with \texttt{Nimbus}. The TwoPop model uses both cloud structures shown in panel I.}
        \label{fig:vhs1256b}
    \end{figure*}
    
    VHS-1256~b is a late L-dwarf (see Table~\ref{tab:planet_param}) which was observed with JWST NIRSpec and MIRI\footnote{Available in MAST: \dataset[10.17909/1563-ws96]{https://doi.org/10.17909/1563-ws96}.} to obtain a full spectrum of VHS-1256~b from 0.97 to 18.02~$\mu$m \citep{miles_jwst_2023}. This spectrum shows clear signs of an Si-O absorption feature. Studies have shown that this Si-O feature can be best explained by a mixture of MgSiO$_3$, Mg$_2$SiO$_4$, and SiO$_2$ \citep{miles_jwst_2023, petrus_jwst_2024}.

    The $T$-$p$, $K_{zz}$, and gas-phase abundances of VHS-1256b were taken from \citet{zhou_roaring_2022} and can be seen in Fig.~\ref{fig:app_all_cloud_structures}. To derive the atmospheric structure, they used the EGP substellar code \citep{marley_atmospheric_1996, fortney_comparative_2005, fortney_planetary_2007, fortney_unified_2008, marley_masses_2012, morley_water_2014, marley_sonora_2021, karalidi_sonora_2021, mukherjee_picaso_2023}. Initial comparisons between the models and the data showed an over prediction of H$_2$O, CH$_4$ and CO$_2$. We therefore reduce the number density of the following species: $\gamma_\mathrm{CH_4} = 0.005$, $\gamma_\mathrm{CO_2} = 0.1$, and $\gamma_\mathrm{H_2S} = 10^{-10}$. The H$_2$O abundances were adjusted for each model individually (see Table~\ref{tab:vhs1256b_tests}). The additional planetary and stellar parameters used for the simulation are listed in Table~\ref{tab:planet_param}.

    \subsubsection{Can \texttt{Virga} explain the observations?}
    \label{sec:vhs_virga}

    We simulate the cloud structure of VHS-1256~b using \texttt{Virga} with a fixed MMR$_\mathrm{base}$ according to Eq.~\ref{eq:chm_mgsio3} (`Basic') and with a free MMR$_\mathrm{base}$ (`Fit MMR'). These runs assume clouds made from only MgSiO$_3$ and a constant $f_\mathrm{sed} = f_\mathrm{base}$. The results are shown in Table~\ref{tab:vhs1256b_tests} and panel G of Fig.~\ref{fig:vhs1256b}. Neither simulation can produce the Si-O absorption feature.
    
    \referee{To solve this problem, we use the TwoPop approach (Sect.~\ref{sec:Model_twopop}).} The $f_\mathrm{sed}$ of the extended cloud follows Eq.~\ref{eq:fsed_param} with $p_0 = 1$~bar, $\alpha = 0.9$, $\beta = 0$, and a variable $f_0$. This parametrisation is necessary to create small high-altitude particles that are not below the physical size limit ($r < 10^{-4}$). The results of the TwoPop cloud profile are shown in Table~\ref{tab:vhs1256b_tests} and panel G of Fig.~\ref{fig:vhs1256b}. The TwoPop approach can reproduce the Si-O feature. 
    
\begin{figure}
        \centering
        \includegraphics[width=1\linewidth]{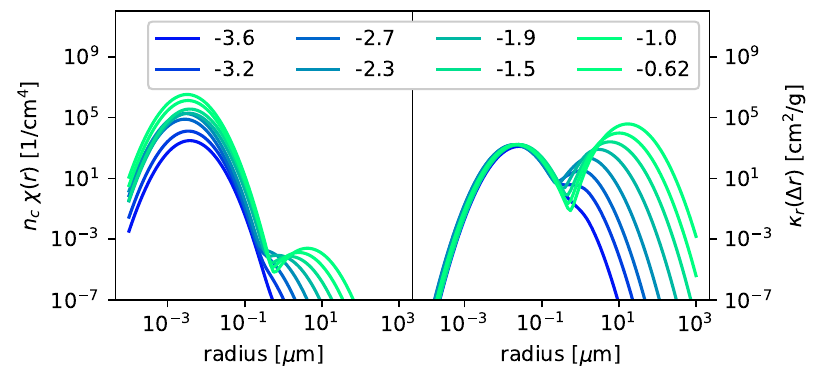}
        \caption{PSD and opacity for the TwoPop approach of VHS-1256~b from panel G of Fig.~\ref{fig:vhs1256b} for different pressure layers (log$_{10}(p)$).}
        \label{fig:vhs1256b_sd}
    \end{figure}

    To analyse the distribution of particles, we plot the PSD of the TwoPop approach in Fig.~\ref{fig:vhs1256b_sd}. While the population of small particles have a larger number density (left panel) throughout the atmosphere, the larger particles have a higher opacity (right panel) near the cloud bottom. The contribution function of the thermal emission spectra confirms that most of the emission originates around 0.1~bar which is close to the pressure layer of the cloud base. However, the Si-O absorption feature originates at pressures where both base and extended cloud have roughly equal opacities.
    
    \subsubsection{Material of the cloud base}
    \label{sec:vhs_material}

    Within the TwoPop approach, the cloud base does not impact the Si-O feature directly. It is therefore possible that the cloud base and the extended cloud are made from different materials. The combination of a Fe cloud with an upper MgSiO$_3$ cloud can in fact explain several observations of brown dwarfs \citep[e.g.][]{burningham_cloud_2021, luna_empirically_2021}. To assess if the cloud base material of VHS-1256~b can be distinguished with our models, we test the following materials: MgSiO$_3$, Mg$_2$SiO$_4$, TiO$_2$, Fe, SiO, SiO$_2$, and Al$_2$O$_3$. The results are shown in Table~\ref{tab:vhs1256b_tests} and panel H of Fig.~\ref{fig:vhs1256b}.

    \referee{A cloud base made from MgSiO$_3$, Mg$_2$SiO$_4$, TiO$_2$, SiO, or Fe all produce a reasonable fit within the limitations of our model set-up ($\Delta\chi^2 < 2$ and $\chi^2_{5.2} < 1$). A cloud base made from , Al$_2$O$_3$, or SiO$_2$ neither leads to a reasonable fit at lower wavelengths ($\Delta\chi^2 > 2$), nor does it reproduce the Si-O feature ($\chi^2_{5.2} > 2$). It is important to note that for these two materials the $f_0$ of the high-altitude MgSiO$_3$ cloud is significantly higher than for all other cases, indicating that these particles compensate for the bad fit of Al$_2$O$_3$ and SiO$_2$ to the cloud base.}

    \subsubsection{Can \texttt{Nimbus} explain the observations?}

    To simulate the cloud structure with \texttt{Nimbus}, we start with a single base cloud made of MgSiO$_3$ (`Base'). We vary the MMR$_\mathrm{base}$ as well as the sticking coefficient $s_\mathrm{base}$. Equivalent to the \texttt{Virga} results, a single \texttt{Nimbus} cloud structure can explain the general shape of the spectrum but cannot produce the Si-O feature. We therefore use the TwoPop approach as well. The extended cloud is made from MgSiO$_3$ as well but has a different sticking coefficient $s_\mathrm{MgSiO_3}$, and MMR$_\mathrm{MgSiO_3}$. All results are shown in Table~\ref{tab:vhs1256b_tests} and panel I of Fig.~\ref{fig:vhs1256b}. The additional small particles from the extended clouds result in a better log-$\chi^2$ value and can reproduce the Si-O feature. \referee{While $s_\mathrm{MgSiO_3} \approx 10^{-10}$ is the best fit, values up to $s_\mathrm{MgSiO_3} \approx 10^{-4}$ produce a nearly identical spectrum. The reason why arbitrarily low growth rates can reproduce the Si-O feature equally well is because the cloud particle approach their lower size limit.}

    \begin{figure}
        \centering
        \includegraphics[width=1\linewidth]{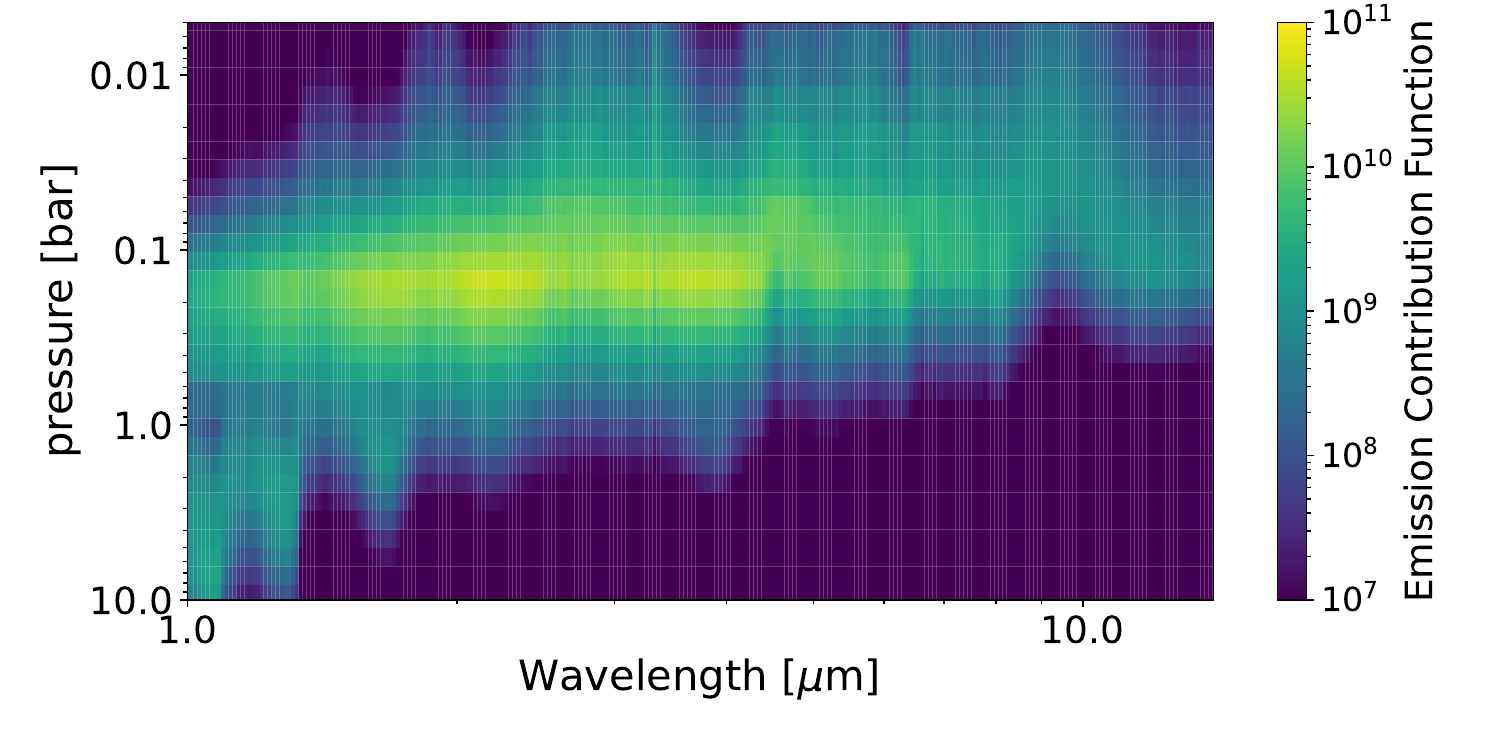}
        \caption{Contribution function for the thermal spectrum of the TwoPop approach for VHS-1256~b from panel G of Fig.~\ref{fig:vhs1256b}.}
        \label{fig:vhs1256b_cf}
    \end{figure}

    \subsection{YSES-1~c}

    \begin{figure*}
        \centering
        \captionof{table}{Parameters for the cloud structure simulations of YSES-1~c with \texttt{Virga} and \texttt{Nimbus}. Boldfaced values are derived through $\chi_\mathrm{log}^2$-minimisation. The following abbreviations are used: 1.23e-4 = $1.23 \cdot 10^{-4}$, MgSiO$_3$ (M3), and Mg$_2$SiO$_4$ (M4). \referee{The $\chi^2_\mathrm{log}$ values are from top to bottom: 2.8e-3, 3.2e-4, 4.3e-4, 4.1e-4, 2.9e-3, 5.1e-4, 3.2e-4, 2.5e-4, 1.6e-4, 4.1e-4, 5.0e-4, and 2.5e-4.}}
        \label{tab:yses1c_tests}
        \begin{tabular}{l l l l l l l l l l l l}
            \hline\hline
             & base & $f_\mathrm{base}$       & MMR$_\mathrm{base}$ & $f_\mathrm{0}$    & MMR$_\mathrm{M3}$ & $\alpha$         & $\gamma_\mathrm{H_2O}$ & $\gamma_\mathrm{CO}$ & $\chi^2$ & $\chi^2_{5.2}$ \\  
            \hline 
            J & M3      & \textbf{1.25}$^{+2.9\textnormal{e}-1}_{-1.9\textnormal{e}-1}$       & 2.59e-3             & -                 & -                 & -                & \textbf{1.00}          & \textbf{1.00}        & 108      & 283          \\
              & M3      & \textbf{0.27}$^{+1.3\textnormal{e}-2}_{-1.2\textnormal{e}-2}$       & \textbf{1.40e-4}$^{+6.9\textnormal{e}-6}_{-6.6\textnormal{e}-6}$    & -                 & -                 & -                & \textbf{1.63}          & \textbf{0.87}        & 12       & 21         \\
              & M4      & \textbf{0.42}$^{+2.7\textnormal{e}-2}_{-2.4\textnormal{e}-2}$       & \textbf{3.77e-4}$^{+4.8\textnormal{e}-5}_{-3.5\textnormal{e}-5}$    & -                 & -                 & -                & \textbf{2.83}          & \textbf{0.41}        & 23      & 49           \\
              & SiO     & \textbf{0.17}$^{+9.4\textnormal{e}-3}_{-8.7\textnormal{e}-3}$       & \textbf{4.26e-5}$^{+2.2\textnormal{e}-6}_{-2.1\textnormal{e}-6}$    & -                 & -                 & -                & \textbf{1.03}          & \textbf{0.42}        & 19       & 42           \\
              & SiO$_2$ & \textbf{1.25}$^{+3.5\textnormal{e}-1}_{-2.6\textnormal{e}-1}$       & \textbf{5.23e-3}$^{+3.1\textnormal{e}-3}_{-8.8\textnormal{e}-4}$    & -                 & -                 & -                & \textbf{1.00}          & \textbf{1.00}        & 65       & 51           \\
              & Fe      & \textbf{0.18}$^{+6.9\textnormal{e}-3}_{-6.5\textnormal{e}-3}$       & \textbf{4.32e-5}$^{+2.2\textnormal{e}-6}_{-2.1\textnormal{e}-6}$    & -                 & -                 & -                & \textbf{1.08}          & \textbf{1.06}        & 28      & 75            \\
            \hline
            K & M3      & \textbf{0.26}$^{+1.3\textnormal{e}-2}_{-1.2\textnormal{e}-2}$       & \textbf{1.35e-4}$^{+6.7\textnormal{e}-6}_{-6.3\textnormal{e}-6}$    & \textbf{2.28e-5}$^{+4.6\textnormal{e}-5}_{-----}$  & \textbf{2.73e-6}$^{+8.1\textnormal{e}-6}_{-----}$  & 0.7              & \textbf{1.63}          & \textbf{1.00}        & 10       & 18         \\
              & M4      & \textbf{0.33}$^{+1.5\textnormal{e}-2}_{-1.3\textnormal{e}-2}$       & \textbf{2.55e-4}$^{+1.7\textnormal{e}-5}_{-1.6\textnormal{e}-5}$    & \textbf{2.05e-5}$^{+9.3\textnormal{e}-6}_{-1.9\textnormal{e}-5}$  & \textbf{2.50e-5}$^{+9.0\textnormal{e}-6}_{+8.6\textnormal{e}-6}$  & 0.7              & \textbf{2.74}          & \textbf{2.09}        & 8.3       & 9.4      \\
              & M4      & \textbf{1.00}$^{+7.7\textnormal{e}-2}_{-6.1\textnormal{e}-2}$       & \textbf{6.10e-4}$^{+7.3\textnormal{e}-5}_{-5.5\textnormal{e}-5}$    & \textbf{3.98e-2}$^{+3.1\textnormal{e}-3}_{-2.8\textnormal{e}-3}$  & \textbf{1.60e-5}$^{+1.6\textnormal{e}-6}_{-1.5\textnormal{e}-6}$  & 0                & \textbf{2.12}          & \textbf{1.00}        & 5.1       & 8.4      \\
            \hline\hline
             & base     & $s_{\mathrm{base}}$ & MMR$_\mathrm{base}$ & $s_{\mathrm{M3}}$ & MMR$_\mathrm{M3}$ & $k_{f}$          & $\gamma_\mathrm{H_2O}$ & $\gamma_\mathrm{CO}$ & $\chi^2$ & $\chi^2_{5.2}$ \\  
            \hline
            L & M3      & \textbf{4.10e-2}$^{+3.9\textnormal{e}-3}_{-3.5\textnormal{e}-3}$    & \textbf{4.53e-3}$^{----}_{-4.4\textnormal{e}-3}$    & -                 & -                 & \textbf{5.23e-2}$^{+4.7\textnormal{e}-3}_{-4.4\textnormal{e}-3}$ & \textbf{3.23}          & \textbf{3.13}        & 25      & 71          \\
              & M4      & \textbf{1.50e-2}$^{+2.3\textnormal{e}-3}_{-2.0\textnormal{e}-3}$    & \textbf{2.21e-5}$^{+8.7\textnormal{e}-6}_{-1.9\textnormal{e}-6}$    & -                 & -                 & \textbf{1.94e-3}$^{+4.9\textnormal{e}-4}_{-5.1\textnormal{e}-4}$ & \textbf{3.83}          & \textbf{3.06}        & 25       & 51            \\
              & M4      & \textbf{9.26e-2}$^{+6.9\textnormal{e}-3}_{-6.4\textnormal{e}-3}$    & \textbf{1.52e-5}$^{+5.4\textnormal{e}-6}_{-3.8\textnormal{e}-6}$    & \textbf{1.08e-4}$^{+1.3\textnormal{e}-4}_{-1.0\textnormal{e}-4}$  & \textbf{2.65e-6}$^{+6.3\textnormal{e}-7}_{-6.8\textnormal{e}-7}$  & \textbf{1.70e-1}$^{+1.5\textnormal{e}-2}_{-1.3\textnormal{e}-2}$ & \textbf{3.84}          & \textbf{10.3}        & 7.5       & 12             \\ 
            \hline
        \end{tabular}

        \vspace{5mm}
        
        \centering
        \includegraphics[width=\linewidth]{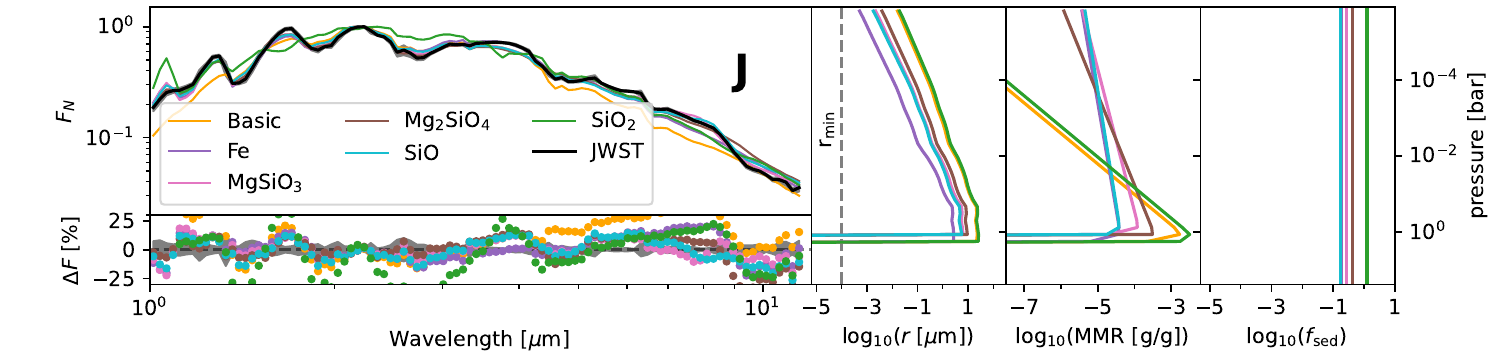}
        \includegraphics[width=\linewidth]{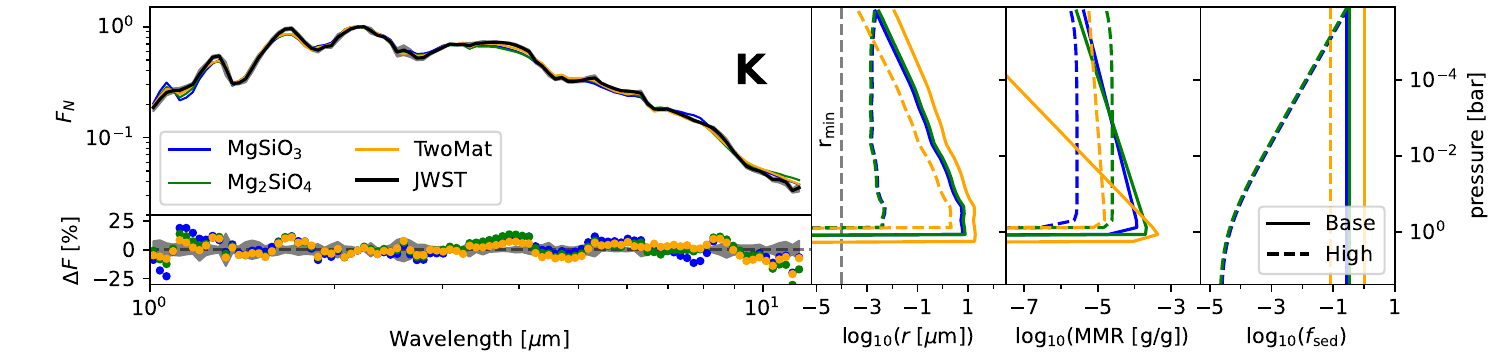}
        \includegraphics[width=\linewidth]{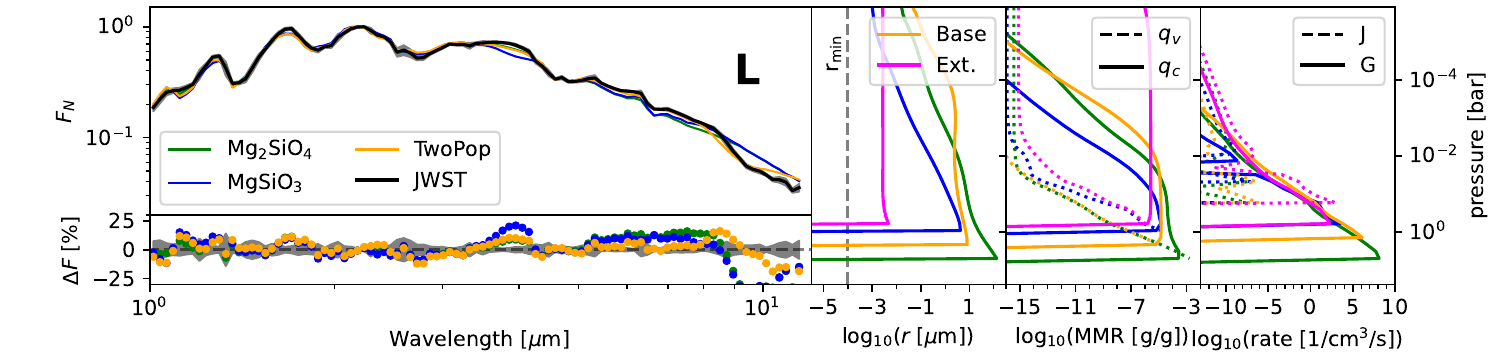}
        \captionof{figure}{Thermal emission spectra and cloud structures models of YSES-1~c. The cloud models of panel J and K are calculated with \texttt{Virga}; the cloud models of panel L are calculated with \texttt{Nimbus}. The `TwoPop' model uses both the `Base' and `Ext.' cloud structures shown in panel L.}
        \label{fig:yses1c}
    \end{figure*}

    YSES-1~c is a young planet around a 17 Myr-old host star (see Table~\ref{tab:planet_param}). JWST observations\footnote{Available in MAST: \dataset[10.17909/1det-0682]{https://doi.org/10.17909/a2vk-mh23}} from 0.6 to 12~$\mu$m revealed a Si-O feature which can be best explained by a combination of MgSiO$_3$ and Mg$_2$SiO$_4$ \citep{hoch_silicate_2025}. Here, we focus on these two materials to model the Si-O absorption feature. The $T$-$p$, $K_{zz}$, and gas-phase abundances of YSES-1c were taken from \citet{hoch_silicate_2025} and can be seen in Fig.~\ref{fig:app_all_cloud_structures}. The H$_2$O and CO abundance are varied for each model separately by scaling the abundances with a factor $\gamma_\mathrm{H_2O}$ and $\gamma_\mathrm{CO}$, respectively. All best fit models are found using a log-$\chi^2$-minimisation (see Eq.~\ref{eq:log_chi_2}). The additional planetary and stellar parameters used for the simulation are listed in Table~\ref{tab:planet_param}.

    \subsubsection{An Si-O feature with \texttt{Virga}}
    
    A `Basic' \texttt{Virga} run assuming MgSiO$_3$ clouds and the MMR$_\mathrm{base}$ listed in Eq~\ref{eq:chm_mgsio3} does not result in a reasonable fit. We therefore use a free MMR$_\mathrm{base}$ and test cloud species with Si-O bonds: MgSiO$_3$, Mg$_2$SiO$_4$, SiO, and SiO$_2$. For comparison, we also use Fe. The results are shown in Table~\ref{tab:yses1c_tests} and panel J of Fig.~\ref{fig:yses1c}. While all Si-O bearing species show the absorption feature, none produce a reasonable fit.

    To achieve a better fit, we use the TwoPop approach. We test cloud bases made from either MgSiO$_3$ and Mg$_2$SiO$_4$ and an extended cloud made from MgSiO$_3$. To achieve the extended MgSiO$_3$ cloud, we use Eq.~\ref{eq:fsed_param} with $p_0 = 1$~bar, $\alpha = 0.7$, $\beta = 0$, and a variable $f_0$. \referee{In the case of two MgSiO$_3$ cloud populations, we find that the high-altitude cloud does not meaningfully contribute. Therefore, it was not possible to determin a lower limit for MMR$_\mathrm{MgSiO_3}$ and $f_0$. We also test a cloud structure (`TwoMat') where both MgSiO$_3$ and Mg$_2$SiO$_4$ have a constant $f_\mathrm{sed}$ throughout the atmosphere ($f_\mathrm{sed} = f_0$ for MgSiO$_3$).} The results are shown in Table~\ref{tab:yses1c_tests} and panel K of Fig.~\ref{fig:yses1c}. We find that in all three cases a combination of both materials is in better agreement with the observations. A combination of MgSiO$_3$ and Mg$_2$SiO$_4$ with a constant $f_\mathrm{sed}$ ($\alpha = 1$) for both clouds leads to the best fit. In all three cases, a clear mismatch in the shape of the Si-O bond feature remains. This is not unexpected as \citet{hoch_silicate_2025} already found that a mixture of materials is necessary to achieve the best fit.

    \subsubsection{Insights from \texttt{Nimbus}}

    We first conduct two \texttt{Nimbus} simulations of YSES-1~c assuming either MgSiO$_3$ and Mg$_2$SiO$_4$ to be the only cloud material. We fit the cloud particle MMR$_\mathrm{base}$, the sticking coefficient $s_\mathrm{base}$, and the diffusion constant $K_{zz}$ (by a factor of $k_f$). The results are shown in Table~\ref{tab:yses1c_tests} and panel M of Fig.~\ref{fig:yses1c}. We find that neither a single MgSiO$_3$ nor a single Mg$_2$SiO$_4$ cloud can produce a Si-O bond feature. \texttt{Nimbus} predicts cloud structures with a steeper decline in cloud particle MMR with altitude than \texttt{Virga}. This leads to a decrease in small particles above the cloud top. We attempted to reduce the growth rate to produce smaller particles. While this did generate the Si-O bond feature, the resulting cloud structure created a clear mismatch at shorter wavelengths and did not yield a good fit.

    To improve the fit, we use the TwoPop approach assuming a base of Mg$_2$SiO$_4$ and an extended cloud made from MgSiO$_3$. The results are shown in Table~\ref{tab:yses1c_tests} and panel M of Fig.~\ref{fig:yses1c}. The TwoPop model achieves a better fit to the spectra than assuming only a single material. However, we can still see clear systematic errors in the residuals between 9 to 11~$\mu$m. This confirms our \texttt{Virga} results that a mix of materials is needed to explain the feature.

\section{Discussion}
\label{sec:Discussion}

    We used \texttt{Virga} and \texttt{Nimbus} to gain insights into the cloud structures of WASP-107b, WASP-17b, VHS-1256~b, and YSES-1~c. With the unprecedented accuracy of JWST we can further constrain exoplanet atmospheres (Sect.~\ref{sec:dis_constraints}). For VHS-1256~b and YSES-1~c, the best match between model and observations was achieved with the TwoPop approached. This cloud structure raises questions on how small particles can reach the upper atmosphere (Sect.~\ref{sec:dis_twopop}). Comparing the cloud structures of the four planets shows how different cloud structures can emerge depending on the atmospheric conditions of the exoplanet (Sect.~\ref{sec:dis_cloud_struct}).

    \subsection{Different cloud structures in exoplanet atmospheres}
    \label{sec:dis_cloud_struct}

\begin{figure*}
        \centering
        \includegraphics[width=\linewidth]{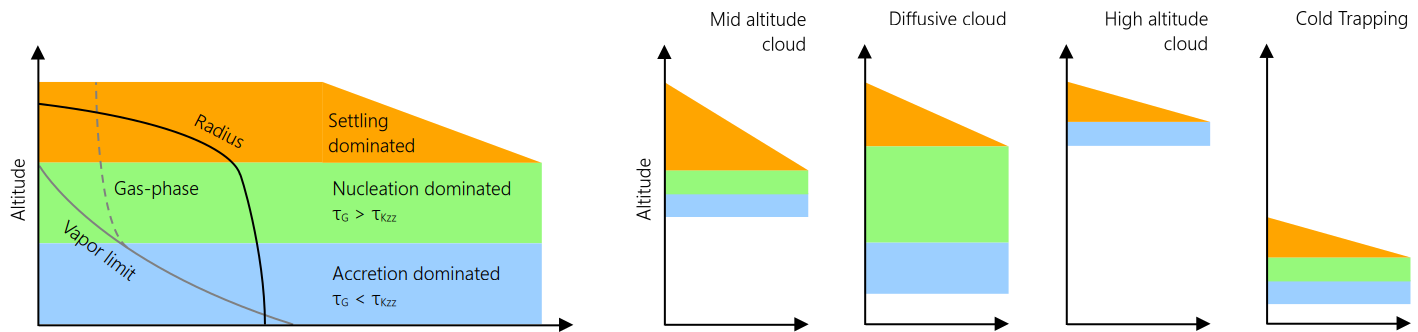}
    
        \vspace{5mm}
    
        \includegraphics[width=\linewidth]{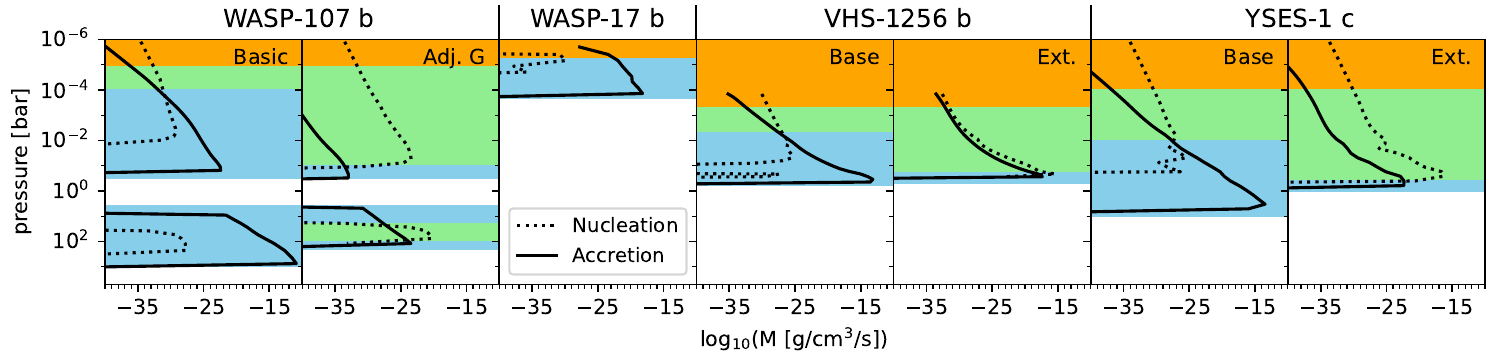}
        \caption{Cloud formation regimes. \textbf{Top:} General mechanism (left) and different cloud structures expected (right). \textbf{Bottom:} Regimes for WASP-107~b (Basic and Adjusted G), WASP-17~b (Base), VHS-1256~b (TwoPop), and YSES-1~c (TwoPop).}
        \label{fig:cloud_struct_illust}
    \end{figure*}

    Each of the four planets analysed in this work share certain underlying cloud structure characteristics. Using the \texttt{Nimbus} results, we can identify three relevant regimes (see Fig.~\ref{fig:cloud_struct_illust}):
    \begin{itemize}
        \item \textbf{Accretion dominated:} The settling cloud particles enable efficient accretion at the cloud base which inhibits nucleation.
        
        \item \textbf{Nucleation dominated:} If growth rates become less efficient than vertical mixing, the partial pressure of the cloud forming species can become significantly larger than the vapour pressure. This allows for nucleation to dominate over accretion.
        
        \item \textbf{Settling dominated:} At lower densities, cloud particles settle more efficiently and only the smallest particles can remain aloft. This defines a `ceiling' for cloud formation.
    \end{itemize}
    Depending on the atmosphere, these three regimes will be at diffent altitudes leading to a range of cloud structures:
    \begin{itemize}
        \item \textbf{Mid altitude clouds} have efficient accretion at the cloud base, nucleation at the cloud top, and form near the settling limit for their particle sizes. 
        
        \item \textbf{Diffusive clouds} occur when cloud particles form where settling is inefficient, allowing them to be mixed upwards. The vertical extent is determined by the mixing strength and growth rate.
        
        \item \textbf{High-altitude clouds} form in hot planets where cloud materials are only stable where settling is efficient.
        
        \item \textbf{Cold trapping} occurs when cloud particles form deep in the atmosphere and are not efficiently mixed upwards. This can be caused by inefficient mixing or efficient accretion.
    \end{itemize}

    \subsubsection{WASP-107~b}
    
    WASP-107~b has a diffusive cloud structure which is surprising considering that Si-bearing species can form efficiently at very low altitudes ($p > 10^{2}$~bar). Similar to the L-T transition in brown dwarfs \citep{suarez_ultracool_2022}, one would have expected that Si is `cold trapped' in deeper layers \citep{powell_formation_2018}. This is what a `Base' \texttt{Nimbus} simulation predicted and what is observed for more refractory species in some hot Jupiters \citep[e.g.][]{parmentier_3d_2013, beatty_evidence_2017, parmentier_transitions_2016, pelletier_vanadium_2023, hoeijmakers_mantis_2024}. If our assumption that clouds form through diffusion of cloud material from the lower atmosphere is correct, accretion has to be significantly less efficient than current theory predicts. Other possible explanations are discussed in Sect.~\ref{sec:dis_twopop}.

    \subsubsection{WASP-17~b}
    
    WASP-17~b is the hottest planet considered in this study and only high-altitude clouds can form. In our models, accretion dominates at higher metallicity, leading to fewer but larger particles. Observations of (ultra)-hot Jupiters like WASP-17~b are ideal to observe ongoing nucleation and accretion to test nucleation models for exoplanet atmospheres \citep[e.g.][]{patzer_dust_1998, lee_dust_2015, lee_dust_2018, bromley_under_2016, gobrecht_bottom-up_2022, gobrecht_bottom-up_2023, sindel_revisiting_2022, lecoq-molinos_vanadium_2024}. However, as our results in Sect.~\ref{sec:WASP-17b} show, it is difficult to constrain cloud formation mechanisms from the Si-O absorption feature alone (see also Sect.~\ref{sec:dis_constraints}).

    \subsubsection{VHS-1256~b and YSES-1~c}
    
    VHS-1256~b and YSES-1~c both have mid-altitude clouds which can explain the general shape of their thermal emission spectra. However, to explain the observed Si-O absorption feature a second extended cloud is required, which resembles a diffusive cloud with a very thin accretion layer. (see Sect.~\ref{sec:dis_twopop}). Our analysis has shown that these two clouds can be either made from different materials, suggesting that different cloud species might have significantly different accretion or settling rates, or be made from the same material, indicating that small and large cloud particles might have different growth rates. Some parametrized models found two material to be a good description of observations, however, their Si-O clouds can be predicted at altitudes above expected rain-out, raising questions on how the cloud particle material reaches the upper atmosphere without condensing.

    \subsubsection{Other studies and limitations}

    Wide-orbit companions may have high-altitude clouds like hot Jupiters, but the 10~$\mu$m feature in thermal emission spectra only probes mid-altitudes (1 to 10$^{-4}$~bars). Similarly, hot Jupiters may have mid-altitude clouds, but transmission spectra are only sensitive to higher altitudes ($p < 10^{-3}$~bar). WASP-107b for example might have optically thick mid-altitude clouds which would be consistent with the TwoPop approach. Combining transmission and thermal emission spectra of hot Jupiters would allow further constraints on the cloud structure.

    VHS-1256~b and YSES-1~c were selected because they show a clear Si-O absorption feature. In cold planetary mass objects and late T dwarfs however, silicate clouds are expected to be below the photosphere, so the spectra are cloud-free \citep{morley_neglected_2012, zalesky_uniform_2022}. In ultra-hot objects, like early L's and ultra-hot Jupiters, clouds of the most refractory species cannot condense at the photosphere \citep{suarez_ultracool_2022, helling_exoplanet_2023}. Furthermore, \citet{suarez_ultracool_2023} found that the absence of Si-O feature in mid-to-late L dwarfs can also be tied to inclination angle, with pole-on geometries. Determining whether the trends found in our study hold for all young, warm exoplanets observed in emission would require a population-level analysis.
    
    In agreement with our work, \citet{burningham_cloud_2021} and \citet{luna_empirically_2021} found that the Si-O feature of brown dwarfs can only be explained with small silicate cloud particles (~0.1 $\mu$m) at low pressures, well above the predicted silicate cloud base. Brown dwarf studies also find that deep clouds made of iron or silicate species are needed to fit near infrared reddening and variability \citep{burningham_cloud_2021, vos_patchy_2023, morley_sonora_2024, mccarthy_multiple_2024, mccarthy_jwst_2025}. The need for two populations of clouds to account for near infrared and mid infrared features is in agreement with our work. 
    
    \subsection{Constraining cloud structures from observations}
    \label{sec:dis_constraints}

    To achieve the best fit, we were required to vary the cloud particle MMR, the settling efficiency, the mixing efficiency, and the growth rate. This allows observational insights into cloud formation processes. However, constraining the physics of cloud formation from observations remains challenging due to degeneracies in observational signatures.  

    \subsubsection{The sticking coefficient \textit{s}}
    
    The sticking coefficient $s$ affects the growth rate and therefore the vertical extent of the cloud. An infinitely thin, optically thick, gray cloud would lead to a blackbody like emission according to the temperature of its atmospheric layer \citep{kattawar_thermal_1970}. In an extended cloud however multiple layers at different temperatures contribute, leading to a broadened emission spectrum. The width of a thermal emission spectrum can therefore be used to determine the vertical extent of the cloud structures  and to constrain the sticking coefficient \citep[see Sect.~\ref{sec:obs_emis} and][]{miles_jwst_2023, petrus_jwst_2024, hoch_silicate_2025}. Our analysis of VHS-1256~b and YSES-1~c indicates sticking coefficients between 0.25 and 0.05 for the base cloud. While experimental values for $s$ of refractory species under exoplanet conditions are missing, our results for the cloud base are in agreement with general expectations from laboratory measurements \citep[$1 > s > 0.01$;][]{vietti_water_1976, palomba_sticking_2001, cail_experimentally_2005, reissaus_sticking_2006, laffon_laboratory-based_2021, furuya_quantifying_2022, bossion_accurate_2024}. 

    \subsubsection{Si-O feature vs panchromatic observations}
    
    The Si-O feature is useful to determine the presence of cloud particles in exoplanet atmospheres and to constrain their composition \citep[e.g.][]{hoch_silicate_2025}. However, our results for WASP-107~b and WASP-17~b show that detections of the Si-O feature alone do not reasonably constrain the cloud structure. All particles with sizes smaller than 1~$\mu$m can produce the feature and the total opacity is degenerate with the cloud particle MMR. Furthermore, applying the TwoPop approach to VHS-1256~b and YSES-1~c has shown that the cloud particles producing the Si-O feature might not follow the same structure as the main cloud deck. Determining the structure of clouds in exoplanet atmospheres therefore requires panchromatic observations. In transmission spectroscopy, a large wavelength coverage allows to combine the Rayleigh slope, the muting of molecular features, and the Si-O bond feature to constrain the cloud particle MMRs, settling efficiencies, and growth rates \citep{welbanks_high_2024, changeat_cloud_2025}. Panchromatic thermal emission spectra allow to derive even better constraints if the photosphere is dominated by clouds (e.g., VHS-1256~b and YSES-1~c). Overall, our results highlight the strength of and the need for panchromatic observations to gain insights into cloud formation processes.

    The shape of the Si-O bond feature can be used to identify the cloud particle materials \citep{wakeford_transmission_2015}. In several retrieval studies, the best fit was achieved using a mix of materials \citep[see e.g.][]{dyrek_so2_2023, inglis_quartz_2024, murphy_panchromatic_2025, changeat_cloud_2025, hoch_silicate_2025}. To calculate heterogenous cloud particles with \texttt{Virga} or \texttt{Nimbus}, multiple simulations with different materials could be conducted. However, this requires an accurate modelling of the gas-phase since materials producing the Si-O bond feature have to compete for silicon \citep{kiefer_fully_2024}. A detailed gas-phase kinetic network in addition to multiple cloud formation reactions is therefore required. The same holds true for the cloud base, which may consist of mixed materials \citep{samra_mineral_2020, kiefer_why_2024} or they might have complex shapes \citep{min_shape_2003, min_modeling_2005, min_absorption_2006, vahidinia_aggregate_2024, lodge_aerosols_2023}

    \subsubsection{The cloud particle MMR}
    
    The cloud particle MMR is linked to the atmospheric metallicity and the strength of vertical mixing. The cloud mass at the base depends on how much condensable material is available, while the amount above it depends on the settling efficiency and decreases with height. The observed cloud particle MMR is therefore a lower limit for the metallicity of the atmosphere. All our best fit models have slightly lower MMRs at the cloud base compared to solar values (Eq.~\ref{eq:chm_sio} to \ref{eq:chm_mg2sio4}). This is in agreement with the findings of other studies for VHS-1256~b \citep{petrus_jwst_2024, lueber_retrieved_2024} and YSES-1~c \citep{hoch_silicate_2025}. For these planets, the difference in MMR compared to solar values is less than a factor of 5. However, allowing a free MMR was crucial to achieve a good fit. Observations of WASP-107~b and WASP-17~b are only sensitive to extended clouds. It was therefore expected that our best fit MMR is lower than other studies who use gas-phase abundances to estimate metallicities \citep{grant_jwst-tst_2023, dyrek_so2_2023, konings_reliability_2025, louie_jwst-tst_2025}. Furthermore, our results suggest that photospheric clouds, which shape the thermal emission spectrum, and high-altitude particles, which create absorption features, might have different compositions (see Sect.~\ref{sec:vhs_material}). It can therefore be challenging for retrieval studies to determine the material of the cloud base since multiple materials can produce a similar fit.  

    \subsubsection{Atmospheric mixing}
    
    Atmospheric mixing shapes the cloud structures by replenishing cloud particle material and lofting cloud particles to higher altitudes. While both strong mixing and inefficient accretion lead to extended cloud structures, we have shown that there is a significant difference in particle radius between these two cases (see WASP-107~b in Sect.~\ref{sec:WASP-107b_nimbus}). While our \texttt{Nimbus} models of YSES-1~c did achieve a better fit with a lower mixing efficiency $k_f$, recovering precise values requires a retrieval framework to study the degeneracies with other cloud formation parameters (e.g., MMR or growth rate)

    \subsubsection{Nucleation rate}
    \label{sec:dis_cons_nucrate}

    \begin{figure}
        \centering
        \includegraphics[width=\linewidth]{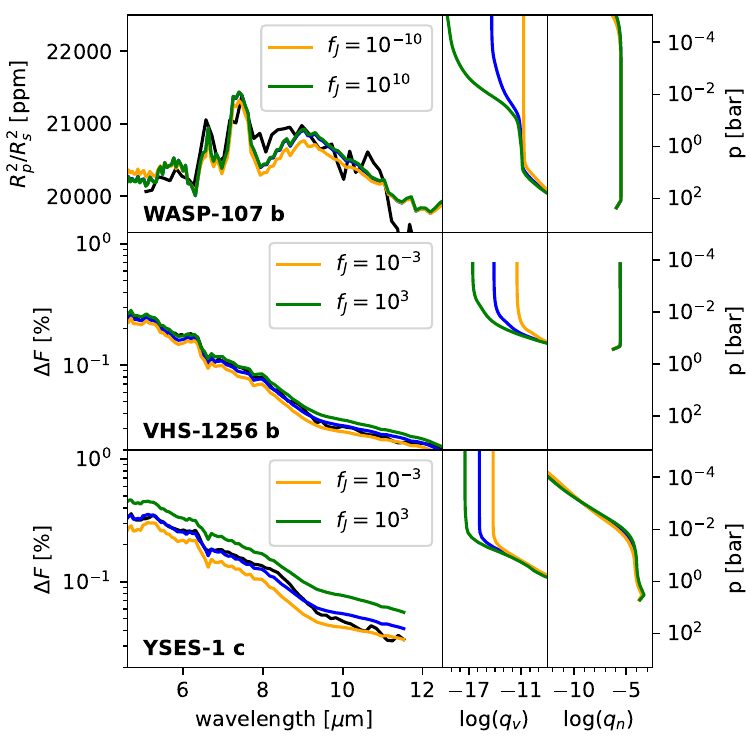}
        \caption{\referee{Impact of nucleation rate changes on the spectrum, gas-phase MMR, and CCN MMR of WASP-107 (Adj.~$G$), VHS-1256~b ('TwoPop'), and YSES~1~c ('TwoPop'). The observations are shown in black and $f_J = 1$ is shown in blue.}}
        \label{fig:nuctest}
    \end{figure}

    \referee{
    Calculating nucleation rates is complex due to the phase transition from gas to solid. While MCNT is an often used approximation \citep[e.g.,][]{helling_dust_2006, gao_microphysics_2018, woitke_dust_2020, lee_modelling_2023}, it has been shown to deviate from non-classical approaches \citep{lee_dust_2015, karthika_review_2016, kohn_dust_2021, kiefer_fully_2024}. To assess if inaccuracies in the nucleation rate matter for our results, we scale the nucleation rate by a factor $f_J$ and analyse the impact on the cloud structure and spectrum. The results are shown in Fig.~\ref{fig:nuctest}. For VHS-1256~b ('TwoPop'), and YSES~1~c ('TwoPop'), increasing or decreasing the nucleation rate within three orders of magnitude ($10^{-3} < f_J < 10^{3}$) results in less than 25\% difference in the spectra. The cloud structure of WASP-107 (Adj.~$G$) is even less sensitive, with ten orders of magnitude change in nucleation rate ($10^{-10} < f_J < 10^{10}$) resulting in less then 400~ppm differences in the spectrum.
    }
    
    \referee{
    The reason for the small difference in the cloud structure despite the drastic changes in the nucleation rate is a self-balancing feedback between nucleation rate and gas-phase abundances in phase-quenched layers. An increased nucleation rate leads to a stronger depletion of the gas-phase material, which in turn decreases the nucleation rate. In all three planets tested, the effect of this feedback can be seen on the gas-phase abundances of the cloud forming materials. To accurately determine the importance of this feedback for substellar atmospheres, a kinetic nucleation study lifting the assumptions of MCNT is required \citep[e.g.,][]{bromley_under_2016, boulangier_devloping_2019, boulangier_developing_2019, kohn_dust_2021, gobrecht_bottom-up_2022, gobrecht_bottom-up_2023, kiefer_fully_2024}.
    }

    \subsubsection{\referee{Current limitations of \texttt{Nimbus}: coagulation}}
    
    \referee{Cloud particle sizes and number densities are affected by collisions with other cloud particles, for example, collisional coagulation, collisional fragmentation, and gravitational coalescence \citep{blum_growth_2008, guttler_outcome_2010}. \citet{ohno_condensation-coalescence_2017} and \citet{samra_mineral_2022} have shown that these processes can alter the cloud structure and affect the optical depth. Cloud-particle-particle collisions are currently neglected in \texttt{Nimbus} to ensure its computational efficiency.
    }
    
    \referee{
    Cloud models for exoplanet atmospheres generally assume that small particles at high-altitudes coagulate, leading to larger but fewer cloud particles \citep{gao_sedimentation_2018, ormel_arcis_2019, ohno_clouds_2020}. It is therefore possible that \texttt{Nimbus} overestimates the cloud particle number densities and underestimated cloud particle sizes. However, since coagulation and gravitational coalescence increase the size of cloud particles, it is unlikely that they present a suitable solution to the origin of the high-altitude silicate particles (Sect.~\ref{sec:dis_twopop}).}

    \subsection{How do small particles reach high altitudes?}
    \label{sec:dis_twopop}

    \subsubsection{Small sticking coefficients}
    \label{sec:dis_twopop_s}
    
    For \texttt{Virga} and \texttt{Nimbus}, it is assumed that clouds form through diffusion of cloud forming materials from the bottom of the atmosphere to the top of the atmosphere. Under these assumptions, small high-altitude cloud particles can only from if $f_\mathrm{sed}$ or the sticking coefficients $s$ are low. Our results show that, to fit observations, the sticking coefficient has to be smaller than $s < 10^{-4}$. 

    Experiments on refractory materials have shown a large range of possible sticking coefficients \citep[$1 > s > 10^{-4}$;][]{palomba_sticking_2001, cail_experimentally_2005, reissaus_sticking_2006}. Some theoretical models predict even lower values \citep[$s < 10^{-4}$;][]{cail_experimentally_2005}. Carbonaceous dust \citep{bossion_accurate_2024}, and water \citep{vietti_water_1976, laffon_laboratory-based_2021, furuya_quantifying_2022} on the other hand typically have larger values ($1 > s > 0.01$). Unfortunately, none of these measurements were conducted under exoplanet atmosphere conditions. Experiments on the sticking coefficient of quartz particles for pressure and temperature ranges of exoplanet atmospheres are therefore necessary to determine whether the observed cloud structures are formed bottom up as modelled by \texttt{Virga} and \texttt{Nimbus}.

    \subsubsection{Efficient nucleation through phase quenching}

    \begin{figure}
        \centering
        \includegraphics[width=1\linewidth]{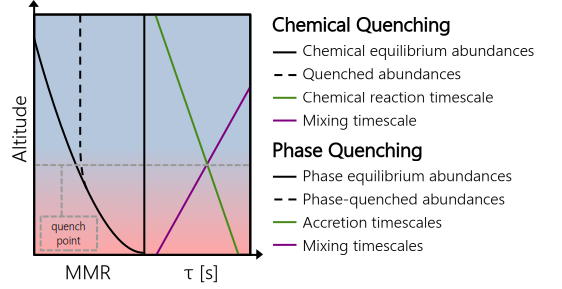}
        \caption{Illustration of chemical and phase quenching.}
        \label{fig:phaseq}
    \end{figure}
    
    Another explanation for the occurrence of small high-altitude particles is active nucleation at high altitudes \citep{molliere_evidence_2025}. Previous theoretical studies have found that nucleation is indeed the dominant cloud formation process in the upper atmosphere \citep{powell_formation_2018, samra_clouds_2022, powell_two-dimensional_2024, kiefer_under_2024} and our results agree with these findings (see Sect.~\ref{sec:dis_cloud_struct}).
    
    To understand in which atmospheric layers nucleation becomes dominant, we introduce the concept of `phase quenching'. This term is derived from chemical quenching in exoplanet atmospheres which occurs when atmospheric mixing is more efficient than chemical reactions, therefore keeping chemical abundances out of equilibrium \citep{moses_chemical_2014}. Similarly, phase-quenching describes atmospheric layers where mixing is more efficient than accretion, leading to a supersaturated atmosphere which is in phase disequilibrium (see Fig.~\ref{fig:phaseq}). Because the nucleation rate is highly sensitive to the amount of supersaturation, phase-quenched layers have increased nucleation. This happens in all simulations with \texttt{Nimbus} and can be recognized by a constant $q_v$. Other cloud models also find phase quenching in the upper atmosphere \citep[e.g.,][]{gao_microphysics_2018, powell_two-dimensional_2024}. While phase quenching can lead to strongly enhanced nucleation rates at high altitudes, it still requires an efficient mass transport of cloud forming material to high altitudes.

    \subsubsection{Atmospheric dynamics}

    Observations of brown dwarfs and exoplanets have shown that clouds can be globally inhomogeneous (`patchy') \citep{miles_jwst_2023, mccarthy_jwst_2025, nasedkin_jwst_2025}. However, one-dimensional models like \texttt{Virga} and \texttt{Nimbus} cannot account for the three-dimensional dynamics of atmospheres. It is therefore possible that the high-altitude silicate particles are a result of global circulation patterns. Previous work has already shown that day to night side circulation can lead to cold trapping of refractory species in hot Jupiters \citep{parmentier_3d_2013}. Local updrafts could, for example, result in an altitude dependent size differentiation of cloud particles because smaller particles have lower settling velocities. These small particles can be advected higher and sustained for longer, potentially explaining the observed population of small high-altitude Si-O particles. However, \texttt{Virga} and \texttt{Nimbus} are not sensitive to this effect because they assume a given PSD in all layers. Detailed studies with binned cloud models are therefore required to study size dependent advection \citep[e.g.][]{lavvas_aerosol_2017, kawashima_theoretical_2018, powell_formation_2018, gao_microphysics_2018}.

    \subsubsection{Meteoritic infall}

    Consistent infall of small meteorites can fuel high-altitude nucleation by providing a source of refractory materials. On Earth this process is studied in connection with the formation of stratospheric aerosols \citep{cziczo_ablation_2001, james_nucleation_2018, james_importance_2023}. In young exoplanetary systems, meteoric impacts might be even more frequent due to the presence of circumstellar or circumplanetary disks \citep[e.g., YSES-1~c and YSES-1~b;][]{hoch_silicate_2025}. Determining if meteorites can explain the observed high-altitude silicate particles requires a dedicated study assessing impact rates in exoplanet systems and nucleation rates of meteoritic materials.

\section{Conclusion}
\label{sec:Conclusion}

    In this work, we demonstrated how a hierarchy of 1D models can be used to investigate cloud structures and the microphysics of cloud formation in exoplanet atmospheres. We investigated WASP-107b, WASP-17b, VHS-1256b, and YSES-1c using \texttt{Virga} and \texttt{Nimbus}. Our analyses of the four planets showed that:
    \begin{itemize}
        \item All four planets have small cloud particles made from Si-bearing species at altitudes higher than the expected cloud base;
        
        \item The sticking coefficient of accretion reactions affects the vertical extent of clouds and can therefore be observationally constrained through thermal emission spectra;
        
        \item The gas-phase abundances of cloud forming materials are `phase quenched`, leading to efficient nucleation at high altitudes;

        \item All four planets are accretion dominated at the cloud base, nucleation dominated above the main cloud deck, and settling dominated at high altitudes.
        
    \end{itemize}
    The best fit cloud structures follow a two-population approach (TwoPop) which combines an optically thick cloud base with an optically thin extended cloud. While our models were able to reproduce observations, the extended clouds have very low $f_\mathrm{sed}$ values and sticking coefficients. Possible explanations for these cloud structures are: 
    \begin{enumerate}
        \item inefficient growth rates of silicon bearing cloud particles due to low sticking coefficients under exoplanet conditions,
        
        \item three-dimensional material transport in combination with efficient nucleation at high altitudes due to phase quenching in combination with, or
        
        \item three-dimensional dynamical effects that trap small clusters in the upper atmosphere, while larger particles gravitationally settle.
    \end{enumerate} 
    
    Panchromatic observations were crucial to achieve constraints on the cloud structures. Ideally, observations should include the Rayleigh slope in the optical ($<1~\mu$m) and the Si-O bond feature in the near infrared ($>9~\mu$m). Our results show that these panchromatic observations have the potential to constrain sticking coefficients $s$, nucleation rates $J$, growth rates $G$ and cloud particle MMRs. By combining microphysical cloud modelling with observations from JWST, or future missions like the Extremely Large Telescope (ELT) and the Habitable Worlds Observatory (HWO), we are able to use exoplanets as extraterrestrial laboratories to study atmospheric chemistry under extreme conditions.

\begin{acknowledgments}
The authors thank David A. Lewis and Vanessa Sennrich for assistance with proofreading and language editing. 
This material is based on work supported by the National Aeronautics and Space Administration under grant No. 80NSSC24M0200 for the NASA XRP program.
Support for program JWST-GO-02044.008-A, JWST-GO-02571.004-A, JWST-GO-01874.007-A, and JWST-GO-05474.006-A was provided by NASA through a grant from the Space Telescope Science Institute, which is operated by the Associations of Universities for Research in Astronomy, Incorporated, under NASA contract NAS5- 26555.
\end{acknowledgments}

\begin{contribution}

SK conducted the simulations and wrote the manuscript. CVM came up with the initial research concept and edited the manuscript. MR provided the observations of YSES-1~c and edited the manuscript.


\end{contribution}

%
\facilities{HST(WFC3), JWST(NIRSpec, MIRI)}

\software{numpy \citep{harris_array_2020},  
          matplotlib \citep{hunter_matplotlib_2007}, 
          xarray \citep{hoyer_xarray_2017, hoyer_xarray_2025},
          SciPy \citep{virtanen_scipy_2020},
          Virga \citep{batalha_condensation_2026, moran_fractal_2025},
          Spellchecking and translation tools (\href{https://www.linguee.com/}{Linguee}, \href{https://languagetool.org}{languagetool}, \href{https://chatgpt.com/}{chatGPT})
          }


\bibliography{references}{}
\bibliographystyle{aasjournalv7}



\end{document}